\documentstyle[12pt]{article}
\def\generic{R$_{1-x}\/$A$_{x}\/$MnO$_{3}\/$\,} 
\def\prsr{Pr$_{1-x}\/$Sr$_{x}\/$MnO$_{3}\/$\,} 
\def\ndsr{Nd$_{1-x}\/$Sr$_{x}\/$MnO$_{3}\/$\,} 
\def\lasr{La$_{1-x}\/$Sr$_{x}\/$MnO$_{3}\/$\,}
\def\lcm{La$_{1-x}\/$Ca$_{x}\/$MnO$_{3}\/$\,}
\def\pcm{Pr$_{1-x}\/$Ca$_{x}\/$MnO$_{3}\/$\,}

\def\tg{$t_{2g}\/$\,} 
\def\eg{$e_{g}\/$\,}

\begin{document}

\begin{center}
{\Large {\bf Magnetic, orbital and charge ordering in the electron-doped manganites
}}\\
\vspace{1.0cm} 
{{\bf Tulika Maitra}$^{*}$\footnote{email: tulika@mpipks-dresden.mpg.de}  
and {\bf A. Taraphder}$^{\dagger}$\footnote{email: arghya@phy.iitkgp.ernet.in, 
arghya@cts.iitkgp.ernet.in}}\\ 

\noindent $^{*}$Max-Planck-Institut f\"ur Physik Komplexer Systeme \\
 N\"othnitzer Str. 38 01187 Dresden, Germany \\
$^{\dagger}$Department of Physics \& Meteorology and  
Centre for Theoretical Studies, \\
Indian Institute of Technology, Kharagpur 721302 India \\   
\end{center}  
\begin{abstract}
\vspace{.5cm} 

The three dimensional perovskite manganites \generic  in the range 
of hole-doping $x > 0.5$ are studied in detail using a double exchange
model with degenerate $e_g$ orbitals including intra- and inter-orbital
correlations and near-neighbour Coulomb repulsion. We show that such a model 
captures the observed phase diagram and orbital-ordering in the intermediate
to large band-width regime. 
It is argued that the Jahn-Teller
effect, considered to be crucial for the region $x<0.5$, does not play a major
role in this region, particularly for systems with moderate to large band-width.
The anisotropic hopping across the degenerate $e_g$ orbitals are crucial in
understanding the ground state phases of this region, an observation emphasized
earlier by Brink and Khomskii. Based on calculations using a realistic limit of 
finite Hund's coupling, we show that the inclusion of interactions stabilizes the 
C-phase, the antiferromagnetic metallic A-phase moves closer to $x=0.5$ while the
ferromagnetic phase shrinks in agreement with recent observations. The charge
ordering close to $x=0.5$ and the effect of reduction of band-width are also
outlined. The effect of disorder and the possibility of inhomogeneous mixture
of competing states have been discussed.  
\end{abstract}
 
\noindent PACS Nos. 75.30.Et, 75.47.Lx, 75.47.Gk

\vspace{.5cm} 
\baselineskip 0.33in 
\noindent {\bf I. Introduction}
\vspace{.5cm}

The colossal magnetoresistive manganites have been investigated with
renewed vigour in the recent past mainly because of their technological
import. It was soon realized that these systems have a rich variety 
of unusual
electronic and magnetic properties involving almost all the known degrees
of freedom in a solid, viz., the charge, spin, orbital and
lattice degrees of freedom\cite{sdm, tokura, coey1}. Of particular interest
have been the systems \generic, where $R$ and $A$ stand for trivalent rare-earth 
(e.g., La, Nd, Pr, Sm) and divalent alkaline-earth ions (Ca, Sr, Ba, Pb etc.) 
respectively. Around the region $0.17 < x < 0.4$, electrical transport 
properties of these systems generically show extreme sensitivity
towards external magnetic field with a concomitant paramagnetic insulator (or
poor metal) to 
ferromagnetic metal transition at fairly high temperatures \cite{cha, RvH, sch}. 
For a long time the dominant paradigm in the theory of this unusual magnetic 
field-dependence of transport has been the idea of $\, double \,\, exchange$
(DE) \cite{zener} involving the localized core spins (the three $t_{2g}$ electrons 
at each Mn site) coupled to the itinerant electrons in the Jahn-Teller split 
\eg level via strong Hund's exchange. It has been realized recently that
such a simplifying theoretical framework may not be adequate in explaining several
other related features involving transport, electronic and magnetic properties
\cite{dagrev1,millis,tvr1,dme,tokura2}. It was already known that the observed 
structural distortions and magnetic and orbital orders in these systems 
in the region $x \simeq 0.5$ require interactions not included in the DE model
\cite{kanamori,good,kugel}. 

Owing to the observation of colossal magnetoresistance (CMR) in the region
$x <  0.5$ in the relatively narrow band-width materials \cite{RvH} at high
temperatures, much of the attention was centred around this region. Only in 
the last few years CMR effect has been observed in the larger band-width
materials like \ndsr\cite{kuwahara95,kuwahara99}  and 
\prsr\cite{tomi95,hejt} in the region $x > 0.5$. If one counts the doping from the 
side $x=1$ in \generic where all Mn ions are in +4 state, then doping by $R_y$ 
($y=1-x$) introduces Mn$^{+3}$ ions carrying one electron in the \eg orbitals.
This region, therefore, is also called the {\it electron-doped region}. The 
charge, magnetic and orbital structures of the manganites in the electron-doped
regime have already been found to be quite rich \cite{tokura,coey2,
tomi96,urusi,imada} and the coupling between all these degrees lead to 
stimulating physics \cite{cuoco}. 
 
In the framework of the conventional DE model with one \eg orbital, one would
expect qualitatively similar physics for $x \sim 0$ and $x\sim 1 $ \cite{emh}. 
On the contrary, experiments reveal a very different and assymetric picture 
for the phase diagram between the regions $x < 0.5 $ and $x > 0.5$.  
The lack of symmetry about $x=0.5$ manifests itself most clearly in the
magnetic phase diagram of these manganites. It has now been shown quite 
distinctly \cite{kuwahara99,kawano97,akimoto98,kajimoto99} that 
the systems \ndsr, \prsr, 
\lasr are antiferromagnetically ordered beyond $x=0.5$ while one observes either
a metallic ferromagnetic state \cite{sch} or a charge ordered state with staggered
charge-ordering \cite{jirac,jirac1} in the approximate range $0.25 < x < 0.5$. This 
charge ordered insulating state can be transformed into a ferromagnetic metallic
state \cite{tomi95,tokunaga98} by the application of magnetic fields.  

There are several different types of AFM phases with their characteristic 
dimensionality of spin ordering observed in this regime. \lasr shows A-type 
antiferromagnetic ground state (in which ferromagnetically aligned $xy$-planes
are coupled antiferromagnetically) in the range $ 0.52 < x < 0.58.$ It also shows 
a sliver of FM phase \cite{akimoto98} immediately above $x=0.5$. In \ndsr
\cite{kajimoto99,kuwahara99}, the 
A-type spin structure appears at $x=0.5$ and is stable upto $x=0.62$ while
in \prsr\cite{tomi95,hejt02}, this region extends from $x=0.48$ to $ x =  0.6$.
In all these cases, 
the phase that abuts the A-type antiferromagnet (AFM) in the region of higher
hole-doping ($x$) is the C-type AFM state, in which antiferromagnetically aligned
planes are coupled ferromagnetically. The C-type AFM phase occupies largest part
of the phase diagram in this region. For even larger $x$, the C-phase gives way to
the three dimensional antiferromagnetic G-phase.  

The systematics of the phase diagram changes considerably (except close to $x=1$)
in these systems as a function of the bandwidth. Recently  Kajimoto et al.
\cite{kajimoto02} have quite succinctly summarized the phase diagrams of various
manganites of varying bandwidths across the entire range of doping. Their phase
diagram is reproduced in fig. 1. The phase diagram changes considerably with 
changing bandwidth as shown in the figure. We note that the narrow bandwidth 
compounds like \pcm, \lcm \,  etc. exhibit a wide region of CE-type insulating 
charge-ordered state around $x=0.5$ whereas the intermediate bandwidth material
\ndsr shows a conducting A-type antiferromagnetic phase around $x=0.5$. As one 
moves towards the larger bandwidth compounds such as \prsr, \lasr, a small
strip of ferromagnetic (F) metallic phase appears at $x=0.5$ \cite{akimoto98,
kajimoto02} followed by the A-type AFM state. In contrast with the narrow 
bandwidth manganites, the relatively wider bandwidth manganites generally show 
the following sequence of spin/charge ordering upon hole doping  (in the 
entire range $ 0 \le x \le  1$): insulating A-type AFM $\rightarrow$ metallic 
FM$\rightarrow$ metallic A-type AFM $\rightarrow$  
insulating C-type AFM and finally insulating G-type AFM states. Clearly, the 
most important feature here is the absence of CE-type spin/charge ordering and 
the presence of a metallic A-type AFM state in these wider band-width compounds
in the region close to $x=0.5$.
It appears that the physics involved in the CE-type charge/spin ordering,
important for the low band-width systems, is not quite as relevant in this case. 
In addition, it is also observed in the neutron diffraction studies that the
metallic A-type AFM state is orbitally ordered \cite{kawano97,kajimoto99}
with predominant occupation of $d_{x^2-y^2}$ orbitals.
The importance of orbital-ordering has been emphasized previously 
in several other experimental \cite{kuwahara99,yoshi98,zimmer01,kawano98} and
theoretical \cite{ishihara,maezono1,maezono2,maezono3,khom97,brink02,long}
investigations. The crucial role of the \eg orbitals and inter-orbital Coulomb
interaction has been underlined by Takahashi and Shiba \cite{takahashi}
from a study of the optical absorption spectra in the ferromagnetic metallic 
phase of the doped manganite \lasr. They point out that it is imperative to 
consider the transition between nearly degenerate and moderately interacting
\eg orbitals even in the hole-doped region in order to interpret the 
optical absorption spectra in \lasr.  

In a detailed observation carried out by Akimoto et al. \cite{akimoto98} 
the electronic and magnetic properties of a heavily doped manganite 
$R_{1-x}Sr_xMnO_3$ with $R=La_{1-z}Nd_z$ are studied by continuously changing 
the band-width. In this novel procedure they were able to control the 
band-width {\it chemically} by changing the average ionic radius by manipulating
the ratio of La and Nd (i.e., changing $z$). Substitution of the smaller 
$Nd^{+3}$ ions for the larger $La^{+3}$ ions effectively reduces the 
one-electron band-width. By increasing $z$ chemically, they were able to go 
continuously from the large band-width system $La_{1-x}Sr_xMnO_3$ down to the 
intermediate band-width system $Nd_{1-x}Sr_xMnO_3$. For $z < 0.5$, there is
a metallic FM phase in the region $0.5 < x < 0.52$. From $x\ge 0.54$ to about
$x=0.58$ the ground state is A-type 
antiferromagnetic metallic irrespective of the value of $z$, i.e. from 
$La_{1-x}Sr_xMnO_3$ ($z=0$) all the way down to $Nd_{1-x}Sr_xMnO_3$ ($z=1$).
They believe that the key factor that stabilizes the A-type AFM metallic state 
in a wide range of $z$ is the structure of the two $e_g$ orbitals ($d_{x^2-y^2}$
and $d_{3z^2-r^2}$) and the anisotropic hopping integral between them. There 
is no signature of charge-ordering or CE-type ordering below $z=0.5$ for any $x$.
The charge-ordered (CO) insulating state appears above $z=0.5$ and around $x=0.5$
primarily due to the commensuration (between the lattice periodicity and hole 
concentration) effect in the low band-width systems. The ground state
phase diagram for doped manganites in $x - z$ plane (i.e., {\it doping versus
band-width} plane) is shown in fig. 2 after Akimoto et al. \cite{akimoto98}. 

The general inferences from all these measurements are that the physics of the
electron-doped region is very different from the hole-doped region. In this region, 
with decreasing band-width starting from \lasr down to \ndsr, the F-phase shrinks,
the A-phase and C-phase remain nearly unaffected. The A-phase disappears and the 
C-phase shrinks (with the possible growth of incommesurate charge order region 
as in fig. 1) rapidly in the low band-width systems like \lcm and \pcm. The G-phase
at the low electron-doping region seems to remain unaffected all through. It has been
seen \cite{akimoto98,kajimoto99,yoshi98} that  
the gradual build up of the AFM correlations in the electron-doped region
is pre-empted by the orbital ordering in the A- and C-phases. The \eg orbitals 
and the anisotropic hopping of electrons between them \cite{kugel, slater}, must
indeed play a significant role given the presence of orbital ordering in much  
of the phase diagram beyond $x=0.5.$ 
It is also realized that the effect of lattice could be ignored in
the first approximation for these moderate to large band-width systems in this
region of doping. All these point to the fact that the interactions that play a
dominant role in the elctron-doped region are different \cite{khom,pai,tmat} 
from the ones that are considered crucial in the hole-doped side. 

There has been a large number of reports of charge ordering and inhomogeneous
states \cite{kuwahara95,kuwahara99,tomi95,wollan,radaelli,zimmer,uehara,cnr} in the 
region $x\simeq 0.5$. These states
are quite abundant in the low band-width materials. The inhomogeneous states
result primarily from the competing ground states \cite{rp} (charge ordered/AFM
and FM in this case) that lead to 1st. order phase transitions with a discontinuity
in the density as the chemical potential is varied. Such transitions are known
to lead to phase separation in the canonical ensemble. Phase separation in this
context has been dicussed in the literature for quite some time \cite{moreo,nagaev,
kagan,guinea1,alonso1,yunoki1}. Such macroscopic phase separations are 
not stable against long range Coulomb interactions and tend to break up into 
microscopic inhomogeneities \cite{nagaev,dagrev2,renner}. There is also the 
well-known CE-type charge and spin ordering that has been seen at $x=0.5$ 
in most of these systems \cite{kanamori,good,wollan} with low band-width. 

In both $Nd_{1-x}Sr_xMnO_3$ and $Pr_{1-x}Sr_xMnO_3$ Kawano et 
al. \cite{kawano97} and Kajimoto et al. \cite {kajimoto02,kajimoto99} have seen 
finite temperature ($T\simeq150K$) first order transitions at $x=0.5$ from a 
ferromagnetic metal to an antiferromagnetic A-phase which is insulating but 
has quite low resistivity (immediately away from $x=0.5$ it becomes A-type 
metal). In a neutron diffraction study Kajimoto et al. \cite{kajimoto02}
have also observed that close to the boundary of the FM and A-type AFM metallic
phases of $Pr_{1-x}Sr_xMnO_3$, an unusual stripe-like charge-order
appears along with this weakly first order transition. 

This stripe-like charge-order is distinctly different from the staggered
charge-ordering of the CE-type state. Very recently, an inhomegeneous
mixture of micron-size
antiferromagnetic grains (possibly charge-ordered) and similar sized ferromagnetic
grains has been seen in electron diffraction and dark-field imaging in the
low band-width system \lcm at $x=0.5$ \cite{loudon} without any evidence of the 
long-range CE or any other macroscopic ordering. The ground state energies of these
different phases seem to be very close \cite{misra} in this region leading to a
possible first order phase transition and consequent phase seggregation. A possible 
nanoscale phase separation between A-type AFM and ferromagnetic regions has 
recently been 
reported by Jirac et al., \cite{jirac2} in the cintered ceramic samples of 
Pr$_{0.44}$Sr$_{56}$MnO$_{3}$ doped with Cr (upto 8 percent). It is also 
observed that although both the ferromagnetic domains and A-type AFM host
are independently metallic (though anisotropic for A-AFM), the resultant 
inhomohgeneous state is non-metallic. 

Almost all the experiments discussed above consider orbital ordering as the 
underlying reason for the various magnetic orders observed in the electron-doped
regime. The anisotropy of the two $e_g$ orbitals and the nature of overlap 
integral between them \cite{kugel,slater} make the electronic bands low dimensional. 
Such anisotropic conduction in turn leads to anisotropic spin exchanges and 
different magnetic structures. In the A-phase the kinetic energy gain of the
electrons is maximum when the orbitals form a 2D band in the $xy$-plane and maximize
the in-plane ferromagnetic exchange interaction. However, in the $z$-direction AFM
super-exchange interaction dominates due to the negligible overlap of $d_{x^2-y^2}$
and $d_{3z^2-r^2}$ orbitals. In addition, the presence of charge ordering and
inhomegeneous or phase separated states, particularly around the commensurate
densities, are suggestive of the vital role of Coulomb interactions in the 
manganites.  
The absence of CE-phase in the moderate to large band-width materials imply
that the role of Jahn-Teller or static lattice distortions may not be as
crucial in the electron-doped regime even in the region close to $x=0.5$.
A model, for the electron-doped systems, therefore, should have as its primary 
ingredients, the two \eg orbitals at each Mn site 
and the anisotropy of hopping between them. In addition, the Coulomb interactions 
are present, and their effects on the charge, orbital and magnetic order
are important \cite{dagrev1,tmat,maezono1,long,misra}. In the next section, we 
motivate a model recently proposed by Brink and Khomskii \cite{khom} for the
electron-doped manganites and later
extended by us \cite{tmat} in order to take into account the effects of local
Coulomb interactions present in these systems. We extend this model further
in the present work, study the magnetic and orbital orders in more detail,  
investigate the possibility of charge-ordering and phase separation and discuss
their consequences. In sections II and III we present
our calculations and results and compare them with experimental literature.
We conclude with a brief discussion on the implications of our results.  
\vspace{.5cm}

\noindent {\bf II.a. Degenerate Double-Exchange Model}
\vspace{.4cm}

Evidently the physics of the region $x > 0.5$ is quite different from 
that in the $x < 0.5$ for the manganites and one has to look at the 
electron-doped manganites from a different perspective. In order to pay
due heed to the compelling experimental and theoretical evidence in support 
of the
vital role of the orbitals, Brink and Khomskii \cite{khom} (hereafter referred
to as BK) have proposed a model for the electron-doped manganites 
that incorporates the $e_g$ orbitals and the anisotropic hopping between them. 
In the undoped LaMnO$_3$ compound each Mn ion has one electron and acts as a 
Jahn-Teller centre, the $e_g$ orbitals are split and the system is orbitally
ordered. Thus for the lightly (hole-) doped system 
one can at the first approxmation ignore the orbital degree of freedom and apply
a single band model like the conventional double exchange (DE) model to describe
it. If, however, one proceeds from the opposite end and starts, for example, from
the insulating CaMnO$_3$ compound where the empty $e_g$ orbitals of Mn$^{+4}$ ions 
are degenerate, then doping trivalent (La, Nd, Pr etc.) ions into CaMnO$_3$ results 
in adding electrons into the doubly degenerate $e_g$ manifold. 

In the doped manganites R$_{1-x}$A$_x$MnO$_3$ there are $y=1-x$ number of 
electrons in the $e_g$ orbitals at each Mn site. Since each site has two
\eg orbitals, four electrons can be accommodated per site and hence the actual
filling (electron density) is $\frac{y}{4}$. At $x=0.5$ (or $y=0.5$), 
corresponding to
the maximum filling in the electron-doped regime, every alternate Mn site has
one electron on the average. This means that the highest filling in the 
electron-doped region is only $\frac{1}{8}$ [we restrict ourselves to the region 
$0.5 \le x \le 1.0$ \, $(0.5 \ge y \ge 0)$ in the foregoing]. Due 
to this low electron concentration and hence very few Jahn-Teller centres 
the \eg band is mostly degenerate and the Jahn-Teller effect is negligible 
to a leading approximation. The neglect of Jahn-Teller effect is also justified
from the experimental evidence presented above. The usual charge and
spin dynamics of the conventional DE model then operate here too, albeit with an
additional degree of freedom coming from the degenerate set of \eg orbitals. 
This process has been described by BK as {\it double exchange via degenerate
orbitals}.

In order to capture the magnetic phases properly, the model has, in addition 
to the usual double exchange term, orbital degeneracy and the superexchange (SE)
coupling between neighbouring $t_{2g}$ spins. At $x=1$ (or $y=0$) end the $e_g$ 
band is 
completely empty and the physics is governed entirely by the antiferromagnetic 
exchange (superexchange) between the $t_{2g}$ spins at neighbouring sites. On 
doping, the band begins to fill up, the kinetic energy of electrons in the 
degenerate $e_g$ levels along with the attendant Hund's coupling between 
$t_{2g}$ and $e_g$ spins begin to compete with the antiferromagnetic 
superexchange interaction 
leading to a rich variety of magnetic and orbital structures. The model used to
describe the ground state properties of the electron-doped manganites contains
the following terms  

$$ H = J_{AF} \sum_{<ij>} {\bf S_{i}}.{\bf S_{j}} - J_{H} \sum_{i} {\bf S_{i}}.
{\bf s_{i}} - \sum_{<ij>\sigma,\alpha,\beta}t_{i,j}^{\alpha \beta} 
c_{i,\alpha,\sigma}^{\dagger} c_{j,\beta,\sigma}\eqno(1)$$

The first term is the usual AF superexchange between $t_{2g}$ spins at nearest-neighbour sites, the second term represents the Hund's exchange coupling between
t$_{2g}$ and $e_g$ spins at each site and the third term stands for the hopping
of electrons between the two orbitals \cite{kugel,slater,and59}
($\alpha,\beta$ take values 1 and 2 for d$_{x^2-y^2}$ and d$_{3z^2-r^2}$
orbitals, corresponding to the choice of the phase $\xi_i=0$ in 
Ref.\cite{hotta00}). The 
hopping matrix elements are determined by the symmetry of $e_g$ orbitals \cite
{kugel,slater}. Although similar in appearance to the conventional DE model the 
presence of orbital degeneracy together with the very anisotropic hopping 
matrix elements $t_{ij}^{\alpha\beta}$ makes this model and its outcome 
very different from the conventional DE model of Zener \cite{zener,emh,furukawa,kubo}
with a single non-degenerate orbital. 

In the manner often used in literature \cite{zener} BK treated the t$_{2g}$
spins quasi-classically and the Hund's coupling was set to infinity. At each
site the spins were allowed to cant in the $xz$-plane leading to the 
effective hopping matrix elements \cite{zener}  $t_{xy}=tcos(\theta_{xy}/2)$
and $t_z=tcos(\theta_z/2).$ Here $\theta_{xy}$ is the angle between nearest 
neighbour t$_{2g}$ spins in the $xy-$plane and $\theta_z$ is the same in the 
$z-$direction. The superexchange energy per state then becomes 
$$E_{SE}=\frac {J_{AF}S_0^2}{2}(2cos\theta_{xy}+cos\theta_z). \eqno(2)$$
In this level of approximation,
the problem reduces to solving the $2\times 2$ matrix equation 
$|| t_{\alpha\beta}-\epsilon\delta_{\alpha\beta} || =0$ for a system of 
spinless fermions. The matrix elements are determined using the standard form
of the overlap integrals \cite{kugel}

$$t_{11}=-2t_{xy}(cosk_x+cosk_y) \eqno(3a)$$ 
$$t_{12}=t_{21}=-\frac{2}{\sqrt{3}}2t_{xy}(cosk_x-cosk_y) \eqno(3b)$$ 
$$t_{22}=-\frac{2}{3}t_{xy}(cosk_x+cosk_y)-\frac{8}{3}t_zcosk_z. \eqno(3c)$$ 

\noindent Here $t_{11}$ is the dispersion due to the overlap between 
$d_{x^2-y^2}$ orbitals on neighbouring sites, $t_{12}$ between a 
$d_{x^2-y^2}$ and a $d_{3z^2-r^2}$ orbital and $t_{22}$ between 
two $d_{3z^2-r^2}$ orbitals. In the foregoing, the system is assumed to
posses a cubic unit cell without any distortion. This is not a serious
drawback, as for these systems in the doping range considered, the deviations
from cubic symmetry are not large \cite{coey2}. Writing $t_{xy}$ and $t_z$ in
terms of $\theta_{xy}$ and $\theta_z$ in the J$_H\rightarrow \infty $ 
approximation, the matrix equation is easily solved to get the energy bands 
$\epsilon_\pm(k)$ as
$$\epsilon_\pm(k)=-\frac{4t_{xy}}{3}(cosk_x+cosk_y)-\frac{4t_z}{3}cosk_z\pm  
([\frac{2t_{xy}}{3}(cosk_x+cosk_y)-\frac{4t_z}{3}cosk_z]^2
+\frac{4t_{xy}^2}{3} (cosk_x-cosk_y)^2)^\frac{1}{2}.\eqno(4)$$

We plot these energy bands in fig. 3 along different symmetry directions 
in the cubic Brillouin zone (BZ) in the A-phase (fig. 3a) and also (fig. 3b)
in the absence of a magnetic order (i.e., for $t_{xy}=t_z=t$) to 
demonstrate the nature of dispersion. In the pure (uncanted) phases the bands
in A- and C- phases become purely two- and one-dimensional. However, we have
plotted the bands for the A-phase in the presence of a small canting in 
fig. 3a. Note that even in the presence of canting, 
there is almost no dispersion in the $z$-direction ($\Gamma$-Z and M-L 
directions). In the canted C-phase as well the band disperses little in the $x$ 
and $y$ directions and remains almost indistinguishable from the pure phase.

The total energy is then obtained for a particular filling by adding the
superexchange contribution to the band energy. It is evident that the
energy spectrum obtained depends on the underlying magnetic structure
as well as the orbital-dependent (anisotropic) hopping matrix elements. 
This will lead to different anisotropic magnetic structures at different
doping. The magnetic phase diagram 
in the (electron) doping $y$ - $t/J_{AF}$ plane is then calculated 
by minimizing the total energy with respect to $\theta_{xy}$ and 
$\theta_z$. The sequence of phases follows from the nature of the DOS 
modulated by the anisotropic overlap of orbitals as well as the DE 
mechanism. At very low doping ($x\sim 1$) BK get a 
stable A-type (canted) antiferromagnetic phase and on increasing the doping 
the system first enters the C-phase and then depending on the value of 
$t/J_{AF}$ directly gets into the ferromagnetic phase or reenters the A-phase 
before becoming ferromagnetic at large doping. The presence of ferromagnetic 
phase at large doping and C-type antiferromagnetic order at the intermediate 
electron doping range is rightly captured in their model. Such a sequence of 
phases is indeed seen in the experimental phase diagram in these systems. Quite
remarkably the phase diagram has almost all the magnetic phases except the 
G-type antiferromagnetic one that is observed experimentally in these systems 
at low electron doping. The phase diagram of BK, unfortunately, has two major 
shortcomings in it. At very low electron-doping a canted A-type 
antiferromagnetic phase is obtained which is stable for all values of $J_{AF}$
whereas 
experimentally G-type antiferromagnetic phase is observed at this end. The 
stability of the G-phase around $x\rightarrow 1$ is quite naturally expected
on physial grounds. At the $y=0 (x=1)$ end there are no electrons in the 
$e_g$ band, the only interaction is the antiferromagnetic exchange between 
neighbouring $t_{2g}$ spins which should lead to the three dimensional G-type 
antiferromagnetic order. The other problem is that of the limiting behaviour.
When the antiferromagnetic exchange interaction is zero or very close to zero 
(i.e. $t/J_{AF}\rightarrow \infty$) the system should be completely ferromagnetic,
a feature which is also missed out in their phase diagram. 

It appears that the designation of the A-type ordering by BK was somewhat 
ambiguous and that might have led to the absence of the G-phase around $x=1$ in 
their phase diagram. This is particularly relevant as 
the typical values of canting obtained by BK in their A-phase are quite large. 
In their convention for different spin ordering, they chose to designate A-
phase when $\theta_{xy} < \theta_z$. It is apparent, therefore, that by
this convention, a spin 
ordering with both the angles $\theta_{xy}$ and $\theta_z$ close to $\pi$ 
but $\theta_{xy} < \theta_z$ , could be designated as a canted A-phase. 
On the other hand, from the structure of spin arrangements, it should be more 
appropriately called a canted G-phase. Although G-phase with such large
canting has not been seen experimentally (there is hardly any evidence
of significant canting in the region close to $x=1$). This ambiguity is easily
resolved if 
in addition one considers orbital ordering which, however, was not included in 
their treatment. We discuss this in more detail later on with reference to our 
calculations. 

The limit of infinite Hund's coupling which BK worked with is unphysical
for the manganites considered \cite{coey1,dagrev1,hotta00,misra}. Typical values
reported in the experiments \cite{coey1,coey2,imada} and various model studies 
\cite{dagrev1,maezono1,hotta00} and LDA calculations \cite{misra,satpat} do
not suggest the spin spilittings of the \eg band in various manganites to be 
very large. These are typically comparable to (or slightly larger than) the
\eg band-width. The scale of
Coulomb correlations are most likely to be even higher \cite{coey1,dagrev1}.
The other serious consequence of using such large values of Hund's coupling
is that the predictions about low energy excitations (like optical spectra, 
specific heat, spin fluctuation energy scales) are going to be inaccurate.
BK's calculation, though, serves as a starting point for improved theories. 

Based on their phase diagram BK argue that the degeneracy of orbitals and 
the anisotropy of hopping are crucial and the lattice (including Jahn-Teller (JT)
effect) is of secondary importance for the physics of electron-doped 
manganites. This was borne out by a more refined calculation by Pai. 
In a more realistic treatment of the spin degrees of freedom, Pai \cite{pai} 
considered the limit of finite J$_H$ in the same model and succeeded in 
recovering the G- and F-phases. 
\vspace{.5cm} 

\noindent {\bf II.b. Double exchange and correlation}
\vspace{.5cm}

We mentioned earlier that by all estimates the Coulomb correlations in 
these systems are large \cite{cuoco,feiner,satpat} and it is 
not obvious, therefore, that the phase diagram obtained by BK will survive once 
these are introduced in the model. Neither of the treatments of BK or Pai 
includes the interactions present in the system, namely the inter- and intra-
orbital Coulomb interactions as well as the longer-range Coulomb interactions. 
Although for low doping the local correlations are expected to be ineffective, 
with increase in doping they preferentially enhance the orbital ordering 
\cite{tmat}. This 
affects the F-phase and alters the relative stability of the A- and C-phases. 
The longer-range part of the interactions would tend to localize the carriers 
and lead to charge ordering. It is, therefore, necessary to include them in 
the Hamiltonian and look for their effects on the phase diagram. In the present
work we have incorporated the onsite inter- and intra-orbital as well as the 
nearest neighbour Coulomb interactions in the model Hamiltonian and studied 
how these terms affect the nature of magnetic phase diagram, orbital ordering 
and other properties of electron doped manganites. We also set out from the 
double exchange model with degenerate $e_g$ orbitals and the superexchange 
interaction between the neighbouring $t_{2g}$ spins. The addition of the 
correlation terms makes the model very different from the ones considered by 
BK and Pai. Besides, the physics of charge ordering is beyond the scope of the 
models earlier considered.

The model Hamiltonian we consider consists of two parts, the first part is the 
same as the Hamiltonian in eqn. (1) we discussed in the previous section. The 
second part, which is the interaction part, has onsite inter- and intra-
orbital interaction and the nearest neighbour Coulomb interaction terms in it. 
The total Hamiltonian is therefore

 $$H=H_1+H_{int}$$
\noindent $H_{1}$ is the same as in eqn. (1) and 

$$H_{int}=U\sum_{i\alpha}\hat{n}_{i\alpha\uparrow}\hat{n}_{i\alpha\downarrow}
+U^{\prime} \sum_{i\sigma\sigma^{\prime}}\hat{n}_{i1\sigma}
\hat{n}_{i2\sigma^{\prime}}+
V \sum_{<ij>}\hat{n}_i\hat{n}_j. \eqno(5)$$

In the above $U$, $U^\prime$ and $V$ are the intra- and inter-orbital and the 
nearest neighbour Coulomb interaction strengths respectively.
We treat the $t_{2g}$ spin subsystem quasi-classically as in BK (this is the 
usual practice in many of the treatments of the double exchange model \cite
{zener,dagrev1,furukawa}), but we choose to work with the more realistic limit of 
finite values of the Hund's coupling. In an uncanted homogeneous ground state 
we choose ${\bf S}={\bf S}_{0} \exp(i{\bf Q.r})$ where the choice of ${\bf Q}$ 
determines the different spin arrangements for the core ( $t_{2g}$ ) spins. 
For example, ${\bf Q}=(0, 0 ,0)$ would be the pure ferromagnetic phase, 
${\bf Q}=(\pi, \pi,
\pi)$ gives the G-type antiferromagnetic phase, ${\bf Q}=(\pi, \pi, 0)$ is for C-
type antiferromagnetic phase and finally ${\bf Q}=(0, 0, \pi)$ reproduces A-type 
antiferromagnetic phase. In the infinite J$_H$ limit, the $e_g$ electron spins
are forced to follow the $t_{2g}$ spins leading to the freezing of their spin 
degrees of 
freedom. At finite J$_H$, however, the quantum nature of the transport allows 
for fluctuations and the $e_g$ spin degrees of freedom, along with anisotropic 
hopping across the two orbitals, play a central role. For canted magnetic 
structures where the angle between two nearest-neighbour $t_{2g}$ spins is
different from
that of the pure phases, ${\bf S}_i$ is given by ${\bf S}_i=S_0(sin\theta_i, 
0, cos\theta_i)$ with $\theta_i$ taking all values between 0 and $\pi$. The
$t_{2g}$ spins are allowed to cant only in the $xz-$plane (this does not cause 
any loss of generality in the treatments that follow). We will discuss the 
canted structures at length in the foregoing. We begin our discussion by 
considering the model without the interaction terms $U$, $U^\prime$  and $V$. 
The interactions and their effects will be dealt with in detail later. 
\vspace{.5cm} 

\noindent {\bf II.c. The non-interacting limit}
\vspace{.5cm} 

Using the usual semi-classical approximation for the $t_{2g}$ spins and the 
choice ${\bf S}={\bf S}_{0} \exp(i{\bf Q.r})$, the Hamiltonian (1) reduces in 
the momentum space to $$ H = \sum_{{\bf k},\alpha,\beta,\sigma} 
\epsilon_{\bf k}^{\alpha\beta} c_{{\bf k}\alpha\sigma}^{\dagger} c_{{\bf k}
\beta\sigma}-J_HS_0 \sum_{{\bf k},\alpha} c_{{\bf k}\alpha\uparrow}^{\dagger}
c_{{\bf k+Q}\alpha\uparrow}+J_HS_0 \sum_{{\bf k},\alpha} c_{{\bf k}\alpha
\downarrow}^{\dagger} c_{{\bf k+Q}\alpha\downarrow}\eqno(6) $$
\noindent where we have followed the notation in \cite{kugel,and59}. 
$\epsilon_{\bf k}^{\alpha \beta}$ are the same as $t^{\alpha\beta}$ introduced 
in eqn. (3)

We can see from the above Hamiltonian that the matrix is now an 
$8\times8$ one with two spins (up and down), two degenerate orbitals 
( $d_{x^2-y^2}$ and $d_{3z^2-r^2}$) with (anisotropic) hopping between them
and two momentum indices 
(${\bf k}$ and ${\bf k+Q}$). Thus, taking a finite value of $J_H$ makes the 
problem $8\times8$ one at each {\bf k}-point in contrast to $2\times2$ 
spinless problem for infinite $J_H$ treated in BK. The superexchange part of 
the ground state energy per state is coming from the first term in $H_1$ and 
is the classical contribution $E_{SE}=\frac{J_{AF}S_0^2}{2}(2cos\theta_{xy}
+ cos\theta_z).$

We diagonalize the Hamiltonian in eqn. (6) at each {\bf k}-point on a finite 
momentum grid. The numerical results converged by a grid size of $64\times64
\times64$. The 
ground state energy is calculated from the eigenvalues for different magnetic 
structures (F, A, C and G) in their uncanted configurations. The magnetic 
structure with minimum ground state energy is determined for each set of 
parameters ($x$, $J_H$ and $J_{AF}$ ) for the entire range of electron doping 
($0.5 \le x \le 1$) to obtain the magnetic phase diagram. In fig. 4 we show the 
ground 
state energies for different magnetic structures for $J_{AF}S_0^2=0.05$ and 
$J_HS_0=16$ around the transition points in the doping range $0.5 \le x \le 1.0.$ 
The value of $J_HS_0$ is chosen somewhat large to compare the figure with that 
when $U^\prime\ne 0$ later. All energies are measured in units of $t$. 
The figure shows that there is a G-type AFM to 
C-type AFM phase transition occuring at $x=0.91$, C-type to A-type transition
at $x = 0.62$ and the A-type AFM to F (ferromagnet) transition at $x = 0.57.$
The procedure is repeated for different values of J$_{AF}$ keeping J$_H$ 
fixed and then reversing the order to generate the full phase diagram.
\vspace{.5cm} 

\noindent {\bf II.d. Magnetic phase diagram and canting}
\vspace{.5cm}

The phase diagram in the $x - J_HS_0$ plane for a typical value of 
$J_{AF}S_0^2=0.05$ is shown in fig. 5. 
There is no general agreement on the values of the parameters involved 
\cite{dagrev1}. From photoemission and optical studies \cite{dagrev1} and 
LDA analysis \cite{satpat} one can glean a range of typical 
values \cite{misra}: $0.1 eV < t < 0.3 eV$, $J_H\simeq 1.5 - 2 
$eV and J$_{AF}\simeq 0.03t - 0.01t$, (Maezono et al. \cite{maezono1,maezono2},
though, quote a lesser value of J$_{AF}=0.01t$, the source of which is a 
possible use of antiferromagnetic transition temperatures in these systems to 
deduce the value of J$_{AF}$). 
We observe that for low values of J$_HS_0$ A-type antiferromagnetic phase is 
stable near $x=0.5$, then C-phase is stabilized for a wide region in the 
intermediate doping range and finally near $x=1$ the G-type antiferromagnetic 
phase has a lower energy. For higher values of $J_HS_0$ the ferromagnetic 
phase has the lowest energy near $x=0.5$ and the sequence of magnetic phases 
from $x=0.5$ to $x=1$ is F $\rightarrow$ A$\rightarrow$  C$\rightarrow$  G. 
All the tranisitions appear to be continuous without any jump in the magnetic
order parameters. 
The general trend observed here is in good accord with the experimental phase 
diagram of the electron-doped manganites of intermediate bandwidth such as 
$Nd_{1-x}Sr_xMnO_3$, $Pr_{1-x}Sr_xMnO_3$ \cite{kajimoto02,kajimoto99,hejt02}. 
Unlike BK we find the G-type antiferromagnetic phase to be stable near $x=1.$ 
This is also in agreemnet with Pai \cite{pai}. 

At the $x=1$ end the degenerate $e_g$ orbitals are completely empty.  
The superexchange interaction is 
isotropic and leads to a three dimensional G-type antiferromagnetic phase. At 
low electron-doping the superexchange still wins over the kinetic energy gain
of the electrons via the development of a ferromagnetic component of spins in
the DE mechanism. Thus the G-phase is stable 
up to a finite electron doping. The value of $x$ where G-phase becomes 
unstable depends weakly on J$_H$ in the experimentally relevant range of J$_H$.
On further increasing the electron doping, the kinetic energy starts 
dominating over the superexchange contribution leading to increased spin 
alignment. This happens because the kinetic energy is an increasing function 
of doping and for small doping it is proportional to the (electron) filling 
whereas the superexchange energy is nearly independent of $x$ \cite{comm_angle}. 
A three-dimensional antiferromagnetic 
spin alignment such as G-phase does not allow the electrons to delocalize for 
the typical values of J$_H$. To gain the kinetic energy the system tries to 
polarize the spins along one, two and finally in all three directions 
successively in the sequence C-, A- and F-phases. The gain in the kinetic 
energy due to such alignments more than offsets the loss in the superexchange 
energy above a particular filling. Thus C-type antiferromagnetic phase with 
ferromagnetically aligned spins along the $z-$direction appears first as we 
increase the electron doping. Then the A-type AFM phase with a two-dimensional 
spin alignment appears and finally the ferromagnetic phase with complete 
alignment of spins is observed. 

The stability of A- and C-phases are further 
enhanced by the ordering of orbitals in these phases. As we show below, the 
A-phase has an orbital ordering of $d_{x^2-y^2}$ type and the C-phase has an 
orbital-ordering of $d_{3z^2-r^2}$ type. The planar $d_{x^2-y^2}$ 
orbital-order in the $xy$-plane in the A-phase and rod-like $d_{3z^2-r^2}$ 
orbital-order in the $z$-direction in the C-phase facilitate the hopping of 
electrons (along the plane for A-phase and across it for the C-phase) with 
a gain in the kinetic energy 
which stabilizes these phases in the respective doping ranges. Hence, it is 
primarily the orbital-order that regulates the DE mechanism  
and leads to the C- and A-type magnetic orders. Such a scenario has been borne 
out in several experiments \cite{akimoto98,kajimoto99,zimmer} where evidence for
orbital ordering is seen at a much higher temperature than the spin ordering. 
However, in the G-and F-phases no significant orbital ordering has been observed.
Thus the interplay of spin alignment along chains or planes and the corresponding 
orbital order leads to the transformation from the one-dimensional to the 
two-dimensional and finally to the three dimensional magnetic structure with 
increased doping. The competition between effective kinetic energy (determined 
by J$_H$, band-filling and orbital-ordering) and superexchange leads to the 
transitions G$\rightarrow$  C$\rightarrow$  A$\rightarrow$  F (with the number
of antiferromagnetic bonds 6, 4, 2 and 0 per site respectively) as the doping 
is varied for a given set of values of $J_H$ and $J_{AF}$. The dimensionality 
of the magnetic and orbital-order in the A- and C-phases described above is 
reflected in the density of state (DOS) in these phases.  
In the A-type AFM phase the dispersion of bands is two-dimensional with a peak 
near the centre of the band and small but nonzero DOS at the band edges when the 
hopping $t_{12}$ between $d_{x^2-y^2}$ and $d_{3z^2-r^2}$ orbitals is zero.
For a finite $t_{12}$ the DOS is still two-dimensional, but the peak at the 
centre splits. In the canted C-phase the DOS is quasi-one-dimensional 
(for $t_{xy} = 0$ it becomes purely one-dimensional) with peaks towards the 
band edges. In the pure C-phase  the band disperses only in the $z$-direction 
and the DOS is one-dimensional. 

Experimentally \cite{akimoto98,kajimoto99} it is observed that there is little
canting in A- and C-phases in most of these systems. This was also emphasized 
by Maezono et. al. 
\cite{maez2} from their theoretical analysis. There are some experimental 
observations \cite{mahen} on $Sm_{1-x}Ca_xMnO_3$ which suggest that the 
G-phase, for low doping, has small canting. Canting of the core spins is 
included in our calculation by writing ${\bf S_{i}}={\bf S}_{0}
(sin\theta_{i},0,cos\theta_{i})$ in the Hamiltonian with $\theta_{i}$ 
taking values between 0 and $\pi$. Such a canted spin 
configuration connects two different spin species (up and down) at the same 
site in contrast with the pure (uncanted) phase.
With this choice of ${\bf S_i}$, the Hund's coupling
term between t$_{2g}$ and e$_g$ spins in the Hamiltonian becomes 
$$ H_{hund}
=-J_HS_0\sum_{i,\alpha}cos\theta_i(c_{i\alpha\uparrow}^{\dagger}c_{i\alpha
\uparrow}-c_{i\alpha\downarrow}^{\dagger}c_{i\alpha\downarrow})-J_HS_0\sum_{i,
\alpha}sin\theta_i(c_{i\alpha\uparrow}^{\dagger}c_{i\alpha\downarrow}+
c_{i\alpha\downarrow}^{\dagger}c_{i\alpha\uparrow}).$$ 

In case of canted magnetic structures the different magnetic phases need to
be defined at the outset. The convention (used by BK as well) to define the
magnetic phases are: The phase is 
A-type when $\theta_{xy} < \theta_z$ as the spins in the $xy$-plane have more
ferromagnetic component than the spins across the planes. Similarly, in the
C-phase $\theta_{xy}$ is taken to be greater than $\theta_z.$ In the canted
G and F phases both $\theta_{xy}$ and $\theta_z$ are close to $180^0$ and 
$0^0$ respectively, although, it is then obvious that the canted G-phase 
and A-phase are synonymous in a certain region \cite{comm_BK}. However, 
orbital order can be used to delineate the two phases.

The ground state energy for different $\theta_{xy}$ and $\theta_z$ is obtained
in exactly the same manner as described above (with U and U$^\prime$ set 
to zero). The qualitative nature of the phase diagram is very similar
to the uncanted phase diagram except for little shifts in the phase 
boundaries (the
shifts are small unless J$_H$ is large). We show in fig. 6a the angle of canting
(for both $\theta_z$ and $\theta_{xy}$) as a function of J$_H$ deep inside the
G-phase at $x=0.98$ (the angles in fig. 6 represent deviation from 180 $^\circ$)
for different values of $J_{AF}S^{2}_{0}$. 

There is hardly any canting in the $z$-direction while in the $xy$-plane there 
is no significant canting for low $J_H$ and it is about 10$^\circ$  only for 
large $J_H$. So, for realistic values of $J_H$, there is no observable canting.
The absence of canting in $\theta_z$ is seen for all the different values of
$J_{AF}S^{2}_{0}$. An increase in $J_{AF}$ reduces canting of $\theta_{xy}$
(fig. 6b) and stabilizes the pure G-phase as expected. Changing $y$ and moving 
closer to the boundary with the C-phase, canting in $\theta_{xy}$ is seen to
increases quite slowly. However, very close to the G-C boundary, $\theta_{xy}$
reverts back towards $\pi$ while $\theta_z$ begins to deviate from $\pi$. 
In the G-phase, small canting has been reported in certain systems as discussed
earlier. For very low electron-doping the superexchange 
interaction wins over double exchange and the phase is G-type antiferromagnetic. 
On doping, the electrons would try to delocalize. Since it is energetically 
costly ($J_H$ being the largest scale) for electrons to move in a completely 
antiferromagnetic configuration it is expected that the system will try to 
gain kinetic energy via the canting of the core spins. The canting angle will 
be anisotropic, i.e.,  $\theta_z$ will be different from $\theta_{xy}$ due to 
the anisotropy of $t_{ij}^{\alpha\beta}$. We should 
also note the fact that canting in the plane leads to higher gain in kinetic 
energy than what is gained by canting in the $z$-direction. This does not, 
however, mean that the phase that abuts G-phase as $x$ decreases would be the 
planar (A-type) magnetic phase - the values of the two angles are delicate 
functions of doping, the dimensionality of the DOS as well as the anisotropy of 
the hopping integral. The phase that appears after G-phase 
with increased doping is the C-phase. In the J$_H$ inifinite limit electron
hopping to neighbouring sites 
with antiparallel core spins is not allowed because the effective hopping 
parameter in this case is proportional to $tcos(\theta/2)$ where $\theta$ is 
the angle between the spins at neighbouring sites and antiparallel arrangement 
of spins reduces it to zero. Hence the only way the electrons can take 
advantage of the kinetic energy gain due to increased doping is by canting the 
core spins as much as possible (at the cost of superexchange energy, which is, 
however, small). This will give rise to a ferromagnetic moment so that the 
electrons can hop from site to site. BK had projected out the ``wrong"  spin 
sector of the Hilbert space in their effective theory with infinite Hund's 
coupling. This is why they observed large canting of spins in the A- and 
C-phases. However, at finite J$_H$ this picture is changed altogether. 
The wrong spins are no longer as  ``costly"  and a finite value of
J$_H$  allows an electron in 
the wrong spin state with an energy cost proportional 
to $J_H$. Hence the canting angle reduces drastically as compared to the 
$J_H\rightarrow\infty$ limit. In fact, for experimentally realistic values of 
$J_H$ the canting is almost negligible as can be inferred from fig. 6. 
It is to be noted that a small canting in the $xy$-plane in the G-phase gives 
rise to a net ferromagnetic correlation in the plane with a value higher than 
that across the layers (which is zero if there is no canting in the 
$z$-direction). Hence one could think of it as a canted A-phase following the 
convention of BK. However, the orbital order, which is present in the A-phase, 
but absent in the G-phase, can be used to distinguish these two phases. 
Moreover, the kinetic energy gain, which, for small doping, is proportional to 
the doping, is not effective in overcoming the SE energy unless $x$ deviates 
from 1 reasonably. Hence one gets a canted G-phase with very small canting 
angles in the region close to $x=1$, resembling the end-member pure G-phase. Since 
the kinetic energy gain is quite small due to the small values of the canting 
angle, this phase does not have any preferential orbital arrangement of the 
$d_{x^2-y^2}$ or $d_{3z^2-r^2}$ type as in the C- and A-phases. Thus the 
stability of the G-phase is primarily due to the dominance of SE energy in the 
region close to $x=1$. This also means that the doping region over which the 
G-phase stabilizes will grow with increase in J$_{AF}$. In particular, for 
J$_{AF}\rightarrow 0$ the system should exhibit ferromagnetism for any doping. 
However, BK find that the phase boundary between the canted G-phase and the 
C-phase does not change significantly as J$_{AF}$ is varied. In contrast, the 
phase diagram we obtained gives a ferromagnetic state for J$_{AF}\rightarrow 0$
for the entire doping regime and the stability region of the G-phase grows with
increase in $J_{AF}$ as expected.  Our results agree in general with the results
of Maezono et. al.\cite{maez2} though the 
A-phase near $x=0.5$ is missing in their work. In a related work, Sheng and Ting 
\cite{sheng} considered the problem from the strong correlation point of view 
in contrast to the band limit that we have adopted. They obtained an effective
model with coupled spin and orbital degrees of freedom in the strong-interaction
limit and use Monte Carlo method to study this model. The C-phase, however,
could not be obtained from their model anywhere in the range $x \ge  0.5$. 
\vspace{0.5cm}

\noindent {\bf III.a. The interacting case: magnetic phases}
\vspace{.5cm}

We treat the three interaction terms in the Hamiltonian (5) in the mean-field 
theory. It has been pointed out by Hotta et al. \cite{hotta00}, that the 
mean-field theory for the 
interacting double-exchange model even in low dimension gives very good 
agreement with exact diagonalization on small systems. Comparison of
mean-field phase diagram with exact diagonalization on small systems by Misra 
et al. \cite{misra} is also quite encouraging. Let us first look into 
the inter-orbital Coulomb interaction term $U^{\prime}\sum_{i\sigma\sigma^
{\prime}}\hat{n}_{i1\sigma}\hat{n}_{i2\sigma^{\prime}}$ and set $U=V=0$ in 
$H_{int}$. In the mean-field theory, one neglects fluctuations and writes  
$ \hat{n}_{i1\sigma}\hat{n}_{i2\sigma^{\prime}}
=<\hat{n}_{1\sigma}>\hat{n}_{i2\sigma^\prime}
+<\hat{n}_{2\sigma^\prime}>\hat{n}_{i1\sigma}-<\hat{n}_{1\sigma}>
<\hat{n}_{2\sigma^\prime}>.$

The homogeneous averages $<\hat{n}_{1\uparrow}>$, $<\hat{n}_{1\downarrow}>$, 
$<\hat{n}_{2\uparrow}>$,
$<\hat{n}_{1\downarrow}>$ were calculated iteratively through successive 
diagonalization of the Hamiltonian. Each of the average quantities and 
the filling were 
calculated from the resultant eigenvectors for a chosen chemical potential and 
fed back to the Hamiltonian for next iteration. All the averages and filling 
were thus allowed to reach self-consistent solutions. Self-consistency is 
achieved when all averages and the ground state energy converge to within 0.01$\%$
or less (depending on the difference in energy with
the competing ground state). In this way the ground state energies are 
calculated at each filling for all four magnetic phases (F, A, C and G) and 
the minimum energy phase was determined to obtain the complete magnetic phase 
diagram in the entire electron-doping regime by varying both J$_H$ and 
J$_{AF}$.

We show in fig. 7 the ground state energies of different magnetic phases around
the transition points with $J_{AF}S_0^2=0.05$, $J_HS_0=16$ and $U^\prime=8$.
We see that the G-C phase transition occurs at $x =0.91$ as before, C-A phase 
transition at $x = 0.57$ and the A-F phase transition at $x = 0.51$. Comparing 
this with fig. 4 we note the shift of position of the transitions. The G-phase
remains uaffected, the C-phase 
widens and F-phase shrinks for $U^{\prime} > 0$. The phase diagram in the $x-
J_H$ plane for $J_{AF}S_0^2=0.05$ and $U^\prime=8$ is shown in fig. 8. The panel
from 8(a)-(c) show the progession of the phase diagram as $U^\prime$ increases.
The $U^{\prime}=0$ phase diagram is shown in 8(a) by dashed lines for comparison. 

It is observed that on increasing $U^{\prime}$ the ferromagnetic phase 
starts shrinking fast, the C-phase gains somewhat while the G-phase remains 
almost unaltered for the entire range of values of $J_{H}S_{0})$ studied. The
trends observed here are in good agreement with the 
experimental observations of Kajimoto et al. \cite{kajimoto02,kajimoto99} 
and Akimoto et al. \cite{akimoto98} (see figs. 1 and 2). The enhanced correlation
effectively reduces the phase space for the electrons. 
The observation \cite{akimoto98,kajimoto02} that on decreasing the band-width, 
the ferromagnetic phase shrinks and finally gets pushed below $x=0.5$ 
with A-phase becoming stable at $x=0.5$ is borne out in fig. 8. The 
stabilities of A- and C-phases are primarily derived from the enhanced 
orbital-ordering in the A- and C-phases driven by the inter-orbital repulsion 
and the low dimensional nature of the DOS.  In the presence of $U^{\prime}$, 
the one-dimensional order leading to the AF instability in the C-phase seems to 
grow faster. Close to the $x=1$ end the electron density is very low, 
there are almost no sites with both the orbitals occupied and $U^{\prime}$
is therefore ineffective. The G-phase remains almost unaffected as in fig. 1.
Similarly the canting of the spin away from $\pi$ observed in the G-phase remains 
the same as in fig. 6. At the other end, however, the electron-density is higher
and the F-phase has preferential occupation of one species of spin at both the 
orbitals. Hence this phase is affected drastically by the inter-orbital repulsion. 

We also compare the phase diagrams with and without $U^{\prime}$ in the 
$x-J_{AF}$ plane for $J_HS_0=10$. The corresponding phase diagram is shown in 
fig. 9. Trends observed in fig. 5 and fig. 8 are also seen in this case. 
The topology of the phase diagram has not changed, though the A-phase 
and F-phase shrink in presence of $U^{\prime}$ while the C-phase has grown.

It is known \cite{hotta00} that at the level of mean-field theory the 
intra-orbital 
repulsion $U$ between opposite spins mimics the effect of $J_H$. As we are 
working with quite low densities (actual filling $\le 0.125$), and the 
relevant $J_H$ values are moderate to large, there is hardly any site with 
both spin species present. Therefore, we find almost no observable effect 
of $U$ on the phase diagram (except for very low $J_H$ where again the 
changes are small) and keep its value zero in the phase diagrams shown.
\vspace{.5cm} 

\noindent {\bf III.b. Magnetic ordering and disorder }
\vspace{.5cm}

The doped manganites \generic are intrinsically disordered owing to the
substitution of trivalent ions by divalent ones. Although the dopant ions
do not enter the active network of MnO$_6$ octahedra that are considered
central to the transport properties and magnetic ordering, their effects 
cannot be ignored. In this kind of substitution not only are the charges on 
the dopant ions different from the trivalent rare-earth ions they replace, 
the ionic sizes of the rare-earths vary considerably (e.g., La, Nd, Pr all 
have different ionic sizes). Hence there is a mismatch of ionic sizes between
these and the divalent ion (like Sr, Ca etc.) that replaces them. Such mismatch
would quite naturally bring about large lattice distortions locally. 

However, the effect of disorder has been completely ignored in the treatments 
discussed so far. BK and Pai \cite{pai} argue that to a first approximation, the
disorder does not seem to play a major role in the magnetic phase diagram 
in this region of doping. This is possibly due to the non-magnetic nature of 
the disorder - the rare earth ions are not found to have any observable moment 
except for Pr and it has been shown that Pr-Mn coupling does not have a 
detectable effect \cite{hwang} in the magnetic structure. The lattice effects
are, in any event, pronounced only close to $x=1.$  

Since the Mn ions are central to the mechanism of magnetic and orbital order 
in the manganites, substitution at this site would be quite revealing.
In the last few years quite a few experimental investigations \cite{ahn,blasco}
have been carried out by substitution of Mn by Fe, Ga, Al. These have similar
ionic sizes and valences as Mn and therefore cause very little distortion in
the lattice \cite{ahn} (though Al-substitution has stability problem beyond
about 10\%). For example, the substitution of Mn$^{+3}$ by Fe$^{+3}$ 
(which has identical ionic size as Mn$^{+3}$  \cite{shan}) in 
La$_{1-x}$Ca$_{x}$Mn$_{1-y}$Fe$_{y}$O$_{3}$ in the AFM region at $x=0.53$
shows that the resistivity increases and magnetoresistance disappears by about
$y=0.13.$ Although the Fe$^{+3}$ has a higher moment than the Mn$^{+3}$ that it 
replaces, one observes a steady suppression of the magnetic moment and 
ferromagnetism with Fe doping \cite{ahn}. Whether there is any accompanying 
changes in the underlying magnetic ordering is not clear. Also the systematics
across several manganites with different band-widths are also not available yet. 

There are two things that happen when Fe is doped in place of Mn: i) In the
octahedral crystal field the Fe$^{+3}$ (high-spin d$^{5}$ configuration) sites
have all their e$_{g\uparrow}$ 
orbitals filled up and hence forbid the motion of electrons from Mn$^{+3}$ into
Fe$^{+3}$ sites thereby preventing DE mechanism to operate and ii) the presence of
an Fe$^{+3}$ instead of 
Mn$^{+3}$ in any site alters the superexchange interaction between this and the
neighbouring sites. It is possible to account for these effects in a qualitative
manner following Alonso et al \cite{alonso2}.

The fraction of Mn$^{+4}$ sites (that is the depletion in the number of electrons
in the system) is increased by $(1-y)^{-1}$ when  $y\ne $ 0 as compared to $y=0$.
For the range of $y$ Ahn et al.\cite{ahn} work with ($ y\sim 0.10 $), this is only
about 10\%. So the effective depletion of electrons and effect (i) can be neglected
to a first approximation deep inside any given phase. Similar situation obtains 
when Al$^{+3}$ or Ga$^{+3}$ (having filled d-band) are doped. 

The change in the SE interaction is approximated by estimating the change in
the effective antiferromagnetic interaction between neighbouring core-spins 
owing to the changed values of them in the coupling of Mn-Mn, Mn-Fe and 
Fe-Fe. The new (effective) J$_{AF}$ is given by

$$J_{AF}^{eff} = J_{AF} [ (1-y)^2 + \frac{5}{3} 2 y (1-y) + \frac{25}{9} y^2] $$

The prefactors (25/9, 5/3 and 1) come from the new spin values involved and 
the factors $(1-y)^2$ etc. are for counting the probability of
sites with Mn-Mn, Mn-Fe and Fe-Fe bonds respectively.
Then, at $y=0.12$ for example, the effective J$_{AF}$ is about 0.06 if the initial
value of J$_{AF}$ is 0.05. This will enhance the AF tendencies (and can even
take the system from F- to A-type AFM phase as in fig. 9 for $x$ close to 0.5)
and increase the resistivity as observed by Ahn et al. 

Although a smaller effect, the depletion of the effective number of electrons 
taking part in the DE mechanism will reduce the conductivity and move the effective
doping $x$ towards right in the phase diagram and increase AF correlations and
resistivity further. There is also the possibility that due to these combined 
effects, the magnetic ground state may get altered, a possibility only further
experiments will reveal.

There is another source of scattering coming from the localized \tg spins at each 
Mn site. The itinerant \eg electrons, in a mean-field sense, can be thought of 
as moving in a magnetic ``field" of the localized spins. It has been shown 
\cite{cerovsky} that such a random field can indeed localize part of the electronic
states, particularly in the low-dimensional bands (as obtain in C- and A-phases). 
Replacing Mn$^{+3}$ by Fe$^{+3}$ with a different moment (5/2 as opposed to 2)
provides random changes in this field and additional channel for scattering. The
observation \cite{dho} of a spin-glass type phase at low temperature in the
Cr-doped La$_{0.46}$Sr$_{0.54}$Mn$_{1-y}$Cr$_{y}$O$_{3}$ ($0< y < 0.08$) is a 
possible indication of how the competing interactions between the coexisting FM
phase in the metallic A-type AFM matrix is affected by scattering off the 
random magnetic Cr-impurity and the resultant localization of mobile  
charge carriers. 
  
\vspace{.5cm} 
\noindent {\bf III.c. Orbital ordering}
\vspace{.5cm}

In the non-interacting case we observed orbital order in both A-phase 
($d_{x^2-y^2}$ type) as well as in the C-phase (of $d_{3z^2-r^2}$ type). 
Such orbital order is also borne out in experiments as discussed above. 
In the interacting
situation, we calculate the orbital occupancies (or the orbital 
density as is customarily called by other workers) from the eigenvectors 
corresponding to the converged ground state solutions for both $d_{x^2-y^2}$ 
and $d_{3z^2-r^2}$ orbitals
in A- and C-phases in their respective regions of stability and 
show the results in fig. 10. As one can see, in the A-phase the $d_{x^2-y^2}$ 
orbital has a higher occupancy whereas in the C phase it is reversed. We check
that the sum of the occupancies of two orbitals is equal to the actual 
filling in both the phases. The three-dimensional magnetically 
ordered F- and G-phases, however, show no orbital ordering, the occupancies in 
both the orbitals are the same. 

The presence of inter-orbital Coulomb interaction $U^{\prime}$ enhances the
orbital ordering in both A- and C-phases as shown in fig. 10. for three different
$U^{\prime}$. Note that at lower electron densities, i.e., as $x$ increases, the
effect of 
$U^{\prime}$ on the orbital occupancies becomes less pronounced and the curves 
for different $U^{\prime}$ merge as expected. We also show the orbital 
occupancies as a function of $U^{\prime}$ in fig. 11a,b in the regions where A- and 
C-phases are stable and the effect of $U^{\prime}$ is noticeable in both the 
A- and C-phases. The orbital densities in C-phase (fig. 11b) attain their 
saturation values by $U^{\prime}\simeq 8$. Since we are interested in the 
region $x\ge0.5$, we have not plotted the orbital densities in A-phase (fig. 11a)
beyond $U^{\prime}=8$ - above this value A-phase shifts below $x=0.5$ at $J_HS_0=5$.
In the large $U^{\prime}$ limit and in the absence of $J_{AF}$ and  $V$, the 
Hamiltonian can be mapped onto a pseudospin Hubbard model \cite{cuoco,yuan} with 
off-diagonal hopping (that breaks the $SU(2)$, while still retaining the global
$U(1)$ symmetry). Such a model overestimates the orbital order \cite{yuan} and
the orbital-paramagnetic state is almost never obtained. 

The orbital order obtained in the A- and C-phases leads to anisotropic
band structures in these phases and this feature becomes sharper as 
$J_H$ increases. In particular, the C-phase has a quasi one-dimensional 
density of states. Ideally, this phase should be conducting in the $z$-direction
along the ferromagnetic chains while insulating in the plane. However,
experimentally one finds this phase to be non-metallic. The nearly 
one-dimensional nature of transport makes it very sensitive to 
disorder, possibly localizing the states. In the A-phase the nature of the occupied 
orbitals impedes electron motion along the $z$-direction, giving rise to a 
large anisotropy between the in-plane and out-of-plane resistivities. 
Therefore the A-phase with its planar
ferromagnetic alignment (and quasi 2D DOS) is not as sensitive to disorder and 
this rationalizes the (in-plane) metallic behaviour in the A-phase seen in 
several experiments \cite{akimoto98,kajimoto02,hejt02} while the C-phase remains 
non-metallic. 
\vspace{.5cm} 

\noindent {\bf III.d. Charge ordering}
\vspace{.5cm}

The nearest-neighbour Coulomb interaction term $V\sum_{<ij>}\hat{n}_i\hat{n}_j$ 
is also 
treated in the mean-field theory with $<\hat{n}_i>=n+C_0exp(i{\bf Q.r}_i)$ where
$C_0$ is
the charge-order parameter and $n$ is the average number of electrons per site
and here ${\bf Q}=(\pi,\pi,\pi)$. We calculate the charge-order parameter $C_0$
self-consistently. A non-zero $C_0$ implies the presence of charge ordering. 
Keeping $U^{\prime}  = 0$, the major change observed in the phase diagram now 
is the absence of the A-phase and the presence of charge ordering for values of
$V > 0.29$. The typical values of $V$ are between 0.2 to 0.5 \cite{coey1,dagrev1}
in units of $t$. Below $V = 0.29$, we do not observe any charge-ordering and 
A-phase reappears. The phase diagrams in the $x-J_HS_0$ plane are 
shown in fig. 12(a)-(c) with $V=0.4$,$V=0.5$ and $V=0.6$ at $U^{\prime}  = 0$. 
Note that there are only 
three phases now. A coexisting charge-ordered ferromagnetic phase, the 
orbitally ordered C-phase and the G-phase. The topology does not change 
appreciably when $U^{\prime}$ is finite. The resultant phase diagram is 
shown in fig. 13. The pattern reflects what is seen in figs. 8 (a)-(c). The
F-CO phase reduces while the C-phase grows slightly with $U^{\prime} $.  

In figs. 12 and 13 a wide region of ferromagnetic charge-ordered (CO) phase 
is observed near $x=0.5$. This observation is in agreement with the recent 
experiments \cite{akimoto98,kajimoto02} where the charge-ordered phase at 
$x=0.5$ is claimed to be 
ferromagnetic in nature (or possibly residing at the boundary of the F- and 
A-phases and straddling both). Although the coexisting F-CO region
that we get is considerably wider than the region observed experimentally.
We do not find any self-consistent solution with both A-phase and charge 
ordering in these phase diagrams for any $V$. 
It is possible that the charge ordering instability is too strong close 
to commensurate ($x=0.5$) filling. The A-phase, being also close 
to $x=0.5$ and deriving its stability from a low-dimensional density of states, 
gets affected by the charge order instability. Both C- and G-phases are 
seen to have no charge ordering in them. The CE-phase at $x=0.5$ seen in the
low band-width systems has a charge stacking along the $z$-direction. Such
a stacking is not favoured by the near-neighbour Coulomb term and the CO state
obtained here has staggered charge ordering in all directions.  

Our observation of the ferromagnetic charge ordered (F-CO) ground state agrees 
qualitatively with the mean-field calculation of Jackeli et. al. \cite{jack}.
They considered a Hamiltonian that has orbital degeneracy, Hund's exchange, 
super-exchange and the near-neighbour Coulomb term and studied the ground state
phase diagram as $V$ and $J_{AF}$ change. There is no local Coulomb 
term in their model. They restrict their calculations to $x=0.5$ and 
$J_H\rightarrow\infty$ limit (using effective hopping integrals) only. 
They obtained charge ordered F, A, C and G-phases 
in the  $J_{AF}S^{2}_{0}\,\, -\,\, V$ plane when the degeneracy of the \eg orbitals 
is neglected. In the degenerate model, the F-CO phase appears only at a critical
value of $V\approx 0.7$. There is no A-phase till $ J_{AF}S^{2}_{0}$ 
reaches 0.1. All the transitions from F-CO phase into AFM states are first order.

We do indeed find a critical value of $V$ for the F-CO phase to appear. The critical
value of $V$ for $J_{H}S_{0}=8$ and $ J_{AF}S^{2}_{0}=0.05$ at $x=0.5$ is about 
0.3, well below the value at $J_{H}\rightarrow \infty$ limit. The larger value 
of critical $V$ is an artefact of the  $J_H\rightarrow\infty$ limit. The tendency
to large canting away from pure AF spin structures is markedly reduced in the 
finite $J_H$ limit as we discussed above. The infinite $J_H$ limit is, 
therefore, expected to overestimate the critical value of near-neighbour
repulsion responsible for CO instability in this model as canting and eventual
ferromagnetic instability with an uniform charge distribution is too strong in
that limit. This critical value is nearly independent of $x$ inside the region
of stability of the F-CO phase for the parameter values we considered. This is
an indication of a possible phase separation (with first order transition)
with part of the system pinned at the commensurate density.  
The CO order parameter $C_0$ has a discontinuous jump at the transition from the 
C-phase into the F-CO phase as shown in fig. 14, which is a signature of a first 
order transition between two states having different magnetic symmetry. 
A similar first order jump has been seen at in previous work \cite{misra,jack}
as well and borne out in several experiments described above. The transition as
a function of $V$ from pure F to F-CO phase appears to be continuous (fig. 15). 
\vspace{.5cm} 

\noindent {\bf IV. Discussion}
\vspace{.5cm}

A summary of the trends observed as a function of near-neighbour interaction
$U^{\prime}$ across the entire range of electron-doping is presented
in fig. 16. A comparison with fig. 1 reveals the similarity between them if
one interprets the increase in $U^{\prime}$ as an effective reduction in the
mobility of electrons and suppression of DE mechanism. The rapid reduction in
the stabilty of F- and A-phases
at large $U^{\prime}$ and an almost unchanged G-phase are indeed observed in fig. 1. 
The C-phase is stable over a wider region of phase diagram in fig. 16 than what is 
experimentally observed.  

There are several appealing features of the model and the limits that we have 
studied in the present investigation. We have been able to show that the phase 
diagram and orbital ordering resemble the experimentally observed ones for the
electron-doped regime to a large degree. By putting in correlations the orbital
orders are enhanced and it was possible to obtain regions of charge-ordering 
close to $x=0.5$. However, there are several interesting questions that 
need to be addressed. The neglect of Jahn-Teller effects may well describe the 
electron-doped manganites in the moderate to large band-width systems and also
works for low band-width systems at low electron-doping. But the presence of CE-type 
ordering at $x=0.5$ in the entire class of low band-width materials remind us 
that the effects are relevant close to this doping. A more complete theory should
account for the Jahn-Teller distorted $Mn^{+3}$ 
sites and evolve from the low band-width to the large band-width description 
successfully. Such a theory, however, is lacking at present. In both
$Nd_{1-x}Sr_xMnO_3$ and $Pr_{1-x}Sr_xMnO_3$ it has been 
seen \cite{hejt02,kawano97,kajimoto02} that at finite temperature ($T\simeq150K$)
there is a first order transition at $x=0.5$ from a 
ferromagnetic metal to an antiferromagnetic A-phase which is insulating but 
has quite low resistivity (immediately away from $x=0.5$ it becomes A-type 
metal). Kajimoto et al. \cite{kajimoto02} have also reported a stripe-like
ordering in the $Pr_{1-x}Sr_xMnO_3$ system coinciding with this weakly first
order transition. There is also the possible phase separation into competing
orders in this region. The model described here does reproduce 
a first order transition from an F-CO state to a C-type AFM state with
concomitant phase separation, albeit with a large region of stability  for 
the CO state. In real systems, with longer range Coulomb interactions present,
the phase seapration is likely to appear as domains of one phase dispersed
in another. Whether this indeed is the mechanism of the inhomogeneous phases
observed or they are intrinsic to the systems \cite{moreo,nagaev,dagrev2,paiprv}
is an open question. Transport properties in this region are going to be intriguing
with possible percolative growth of FM clusters in an applied magnetic field
as an alternate route to negative magnetoresistance as opposed to the DE mechanism.  

Extending the model we considered with the possible inclusion of lattice 
degrees of freedom and from a finite temperature calculation, it should be 
possible to look into stripe formations and anisotropic charge orders. It has
been suggested \cite{min} recently
from a finite temperature mean-field calculation with a degenerate, 
non-interacting DE model at infinite $J_H$ limit that without the Jahn-Teller
physics brought in, the CE-phase at $x=0.5$ in the low band-width system is 
not accessible. Though the possibility is wide open \cite{brink02} in the presence 
of Coulomb interactions like $U^{\prime}$ and $V.$  

There is a major class of layered manganites for which the electron-doped
side is still unexplored in detail. The bi-layer systems like 
La$_{2+2x}$Sr$_{1-2x}$Mn$_{2}$O$_{7}$ have shown \cite{lin} similar anisotropic 
magnetic structures as in 3D manganites. Preliminary results from a 
mean-field analysis \cite{tmat2} show interesting promise. It is difficult though 
to account for the large region of C-type ordering seen in experiments in such
layered systems at doping ($x$) ranges as high as 0.75-0.90. In the layered systems 
the DE mechanism is expected to favour either a planar A-type or a G-type state, 
depending on the carrier concentration, over the 1D C-like ordering. The present
model may need additional inputs in order to understand the layered manganites. 

We have not looked into the excitation spectrum of the manganites so far. The
effects of fluctuation coming from both spin and orbital degree and their
coupling may lead to complicated excitations \cite{cuoco,ishihara,caste}. They 
will affect the thermodynamics quite strongly. The controlled incorporation
of disorder, particularly without affecting the lattice \cite{ahn,blasco},
has opened up a host of possibilities. The observation of non-metallic
behaviour in an inhomogeneous mixture of two metallic phases \cite{jirac2}
is an indication of the complex nature of coupling across the boundary of such
domains. The spin-glass like phase reported close to the border of hole- and
electron-doped region \cite{dho} in  La$_{0.46}$Sr$_{0.54}$Mn$_{1-y}$Cr$_{y}$O$_{3}$ 
is another manifestation of the complicated coupling of the impurity with spin 
and charge degrees of freedom. More results of such impurity doping in the 
electron-doped manganites are expected in the near future. We have extended
the model we used to incorporate some of these effects \cite{tmat3} and it would
be quite instructive to investigate the nature of coupling between the impurity
and the magnetic and orbital degrees of freedom.  
\vspace{0.5cm}

\noindent {\bf Acknowledgement}

The research of AT has been funded by a grant from the department of scince and
technology, govt. of India. We acknowldge several useful discussions with S. D.
Mahanti and G. V. Pai. Discussions with S. K. Ghatak and Rahul Pandit are also
acknowledged.

\newpage
\center {\Large \bf Figure captions}
\vspace{0.5cm}
\begin{itemize}

\item[Fig. 1] Schematic phase diagram in the band-width versus hole concentration
in the series of three dimensional manganites after Kajimoto et al. 
\cite{kajimoto02}. The labels represent different magnetic phases explained
in the text. C$_x$E$_{1-x}$ stands for an incommensurate charge-ordered and
CE-type spin ordered phase.
 
\item[Fig. 2] Phase diagram in $z$ versus hole concentration plane for 
(La$_{1-z}$Nd$_z$)$_{1-x}$Sr$_x$MnO$_3$ after Akimoto et al.\cite{akimoto98}.
The effective band-width decreses as $z$ increases. COI stands for charge-ordered
insulating phase. 

\item[Fig. 3.] (a) Band dispersions in the A-phase along the different symmetry
directions of a cubic Brillouin zone. Note the lack of dispersion along
$z$-direction. In (b) is shown the dispersion in the magnetically isotropic state  
where the upper band now disperses along $\Gamma-z$ direction.

\item[Fig. 4.] Ground state energy of different magnetic phases versus
hole-concentration $x  > 0.5$
close to the respective transitions (F-phase to A-phase in (a), A-C in (b) and
C-G in (c)) for $J_{H}S_{0}=16$ and  $J_{AF}S^{2}_{0}=0.05$. All energies are measured
in units of hopping $t$.

\item[Fig. 5.] Magnetic phase diagram in doping ($x$) - J$_H$S$_{0}$ plane with 
$U^\prime=0.$ 

\item[Fig. 6.] Canting of the angles $\theta_{xy}$ and $\theta{z}$ in degrees
(a) as a function of J$_{H}S_0$ for J$_{AF}S_{0}^{2} = 0.04$ (solid line),$0.05$ (dotted line) and $0.06$ (dashed line)  
and (b) $\theta_{xy}$ versus J$_{AF}S_{0}^{2}$ at J$_{H}S_0=10$.

\item[Fig. 7.] Same as Fig. 4, in the presence of onsite inter-orbital Coulomb
interaction $U^{\prime}$.  

\item[Fig. 8.] Magnetic phase diagram in doping ($x$) - J$_H$S$_{0}$ plane
for different $U^{\prime}$. Note the gradual shrinking of the F-phase in
the region  $x > 0.5$. For low $U^{\prime}$ the size of the A-phase remains  
unaffected but at larger $U^{\prime}$ it rapidly shrinks. The C-phase
grows a bit while the G-phase remains nearly unaffected.

\item[Fig. 9.] Magnetic phase diagram in doping ($x$) - J$_{AF}$S$^{2}_{0}$ plane 
with $U^\prime=0$ and 8. 

\item[Fig. 10.] Orbital densities as a function of doping $x$ for three values
of $U^\prime=0, 4, 8$. The filled symbols are for $d_{z^2}$ and open symbols for
$d_{x^2-y^2}$ orbitals. The vertical dotted lines represent the boundary
between A- and C-phases for different $U^\prime$. We choose J$_{H}S_{0}=5$
here in order to have stable A- and C-phases for a reasonable range of $x$
(see figs. 5 and 8) for all three $U^\prime$ values. J$_{AF}S_{0}^{2}$ was 
kept at 0.05.

\item[Fig. 11.] Orbital density versus $U^\prime$ in (a) A-phase at $x=0.5$
and (b) C-phase at $x=0.65$. The dotted lines are for $d_{z^2}$ and solid
lines are for $d_{x^2-y^2}$ orbitals. J$_{H}S_{0}$ and J$_{AF}S_{0}^{2}$ 
were same as in fig. 10.

\item[Fig. 12.]  Magnetic phase diagram in doping ($x$) - J$_H$S$_{0}$ plane
for three different $V$. The F-CO region gets wider with increasing $V$. 
The F-CO to C transition is first order and the inhomogeneous boundary region
is shown with shading. The other transitions are continuos as in figs. 5 and 8. 

\item[Fig. 13.]  Magnetic phase diagram in doping ($x$) - J$_H$S$_{0}$ plane
for finite $V$ at two different values of $U^\prime$. Note that on changing
$U^\prime$ the trend follows that in figs. 5 and 8. The F-CO to C transition
is not shaded here to show the effect of changing $U^{\prime}$.  

\item[Fig. 14.] The charge-order parameter versus hole concentration for
J$_{AF}S_{0}^{2}=0.05$.  

\item[Fig. 15.] The charge-order parameter versus near-neighbour Coulomb
interaction strength for two different hole concentrations. The transition
F to F-CO as a function of $V$ is continuous.

\item[Fig. 16.] Summary of the general trend observed  in  the various
phase diagrams (for $V=0$). Note the trend with increasing $U^{\prime}$
follows closely that of fig. 1 with decreasing band-width.  
\end{itemize}

\begin{thebibliography}{999}
\bibitem{comm_BK} The phase identified as canted A-phase by BK is actually  
a canted G-phase. This phase should be contrasted with the A-phase close
to $x=0.5$ end which is orbitally ordered.
\bibitem {comm_angle}  The changes in $\theta_{xy,z}$ for small electron-doping
are small and depend weakly on $y$. Furthermore, the change in SE energy depends
quadratically on the changes in $\theta_{xy,z}.$
\bibitem{f2} In the choice of J$_H$S$_0$ in eqn.(3), there is an additional 
factor of 2 in ref.\cite{pai}.
\end{thebibliography}
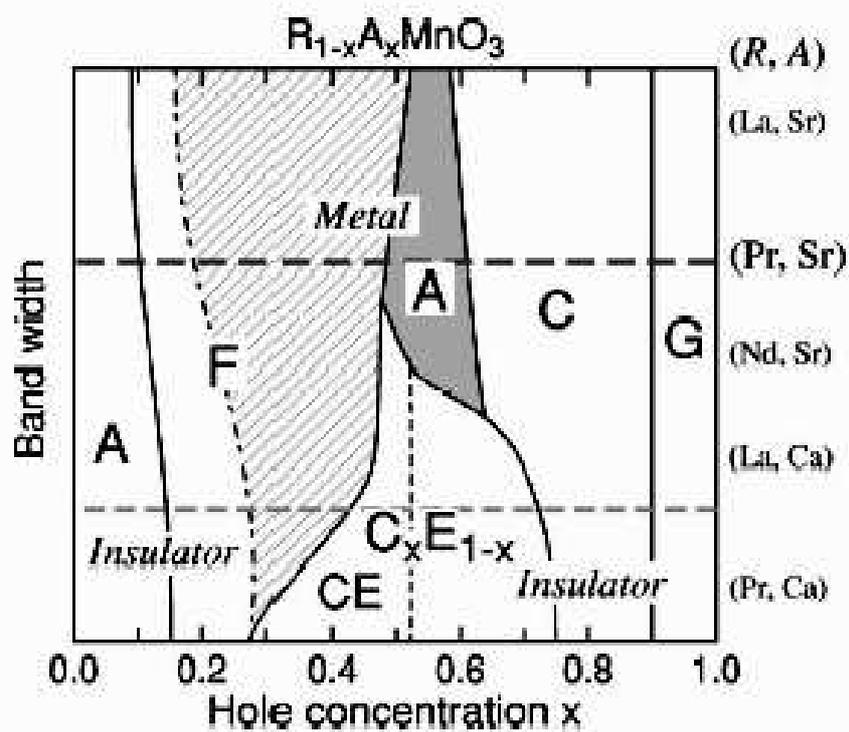
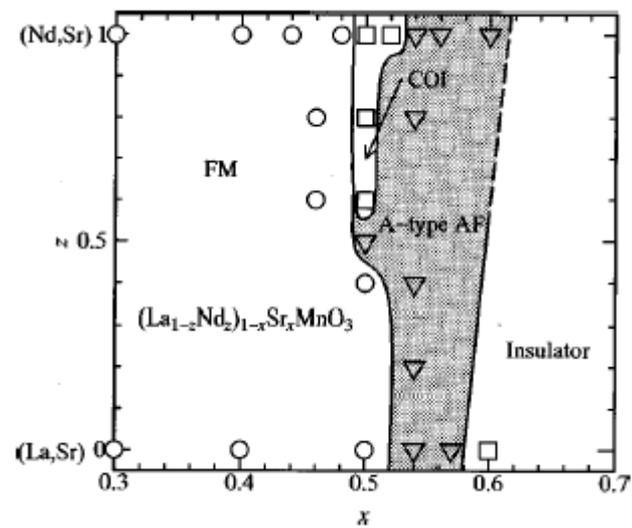
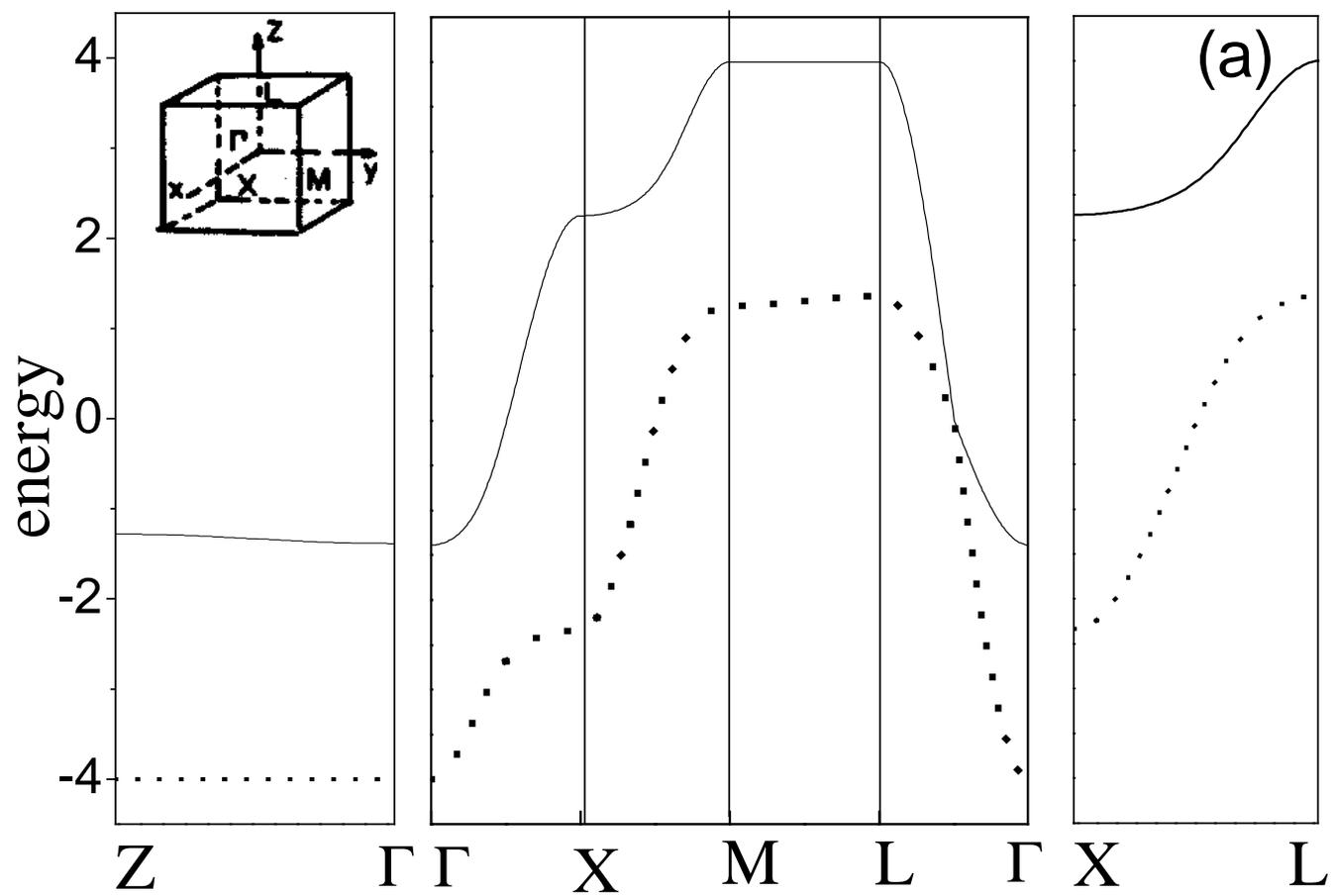
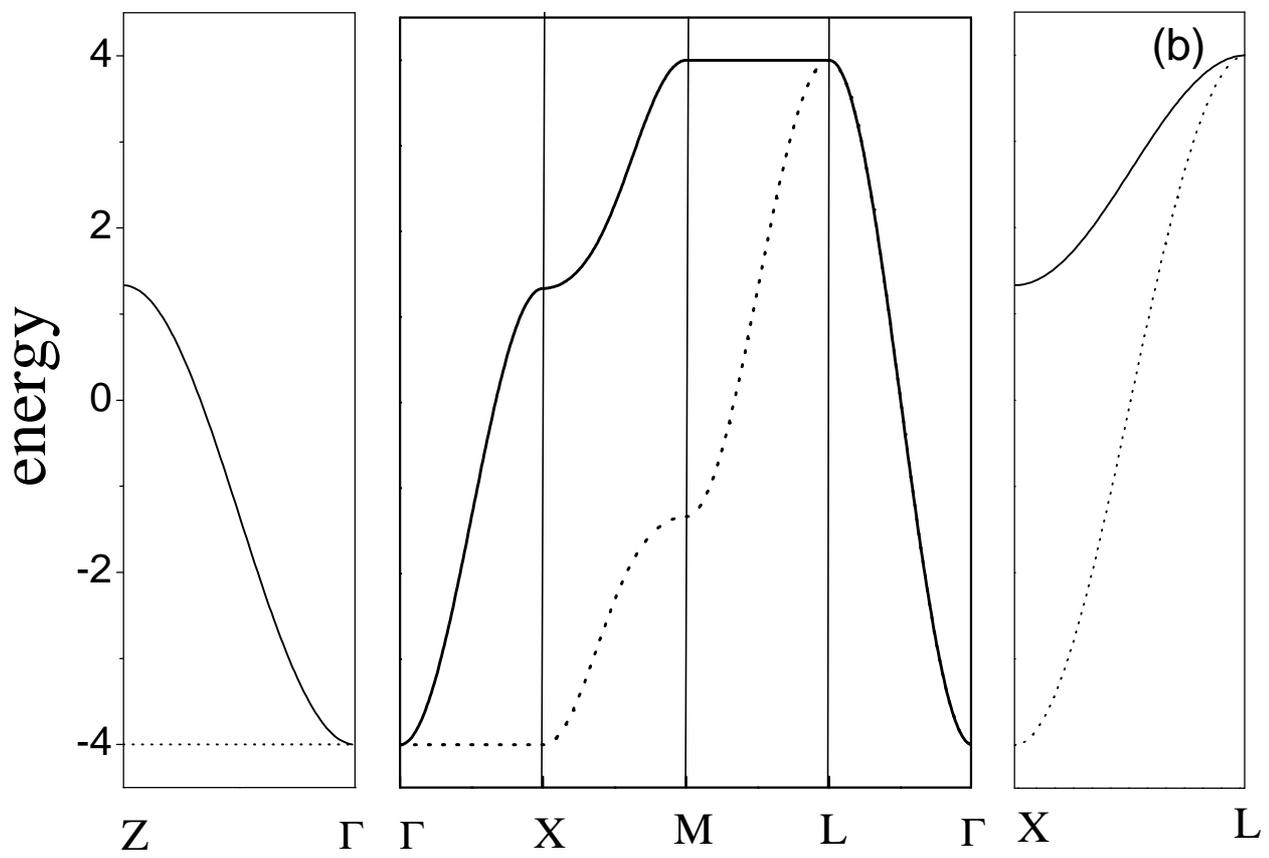
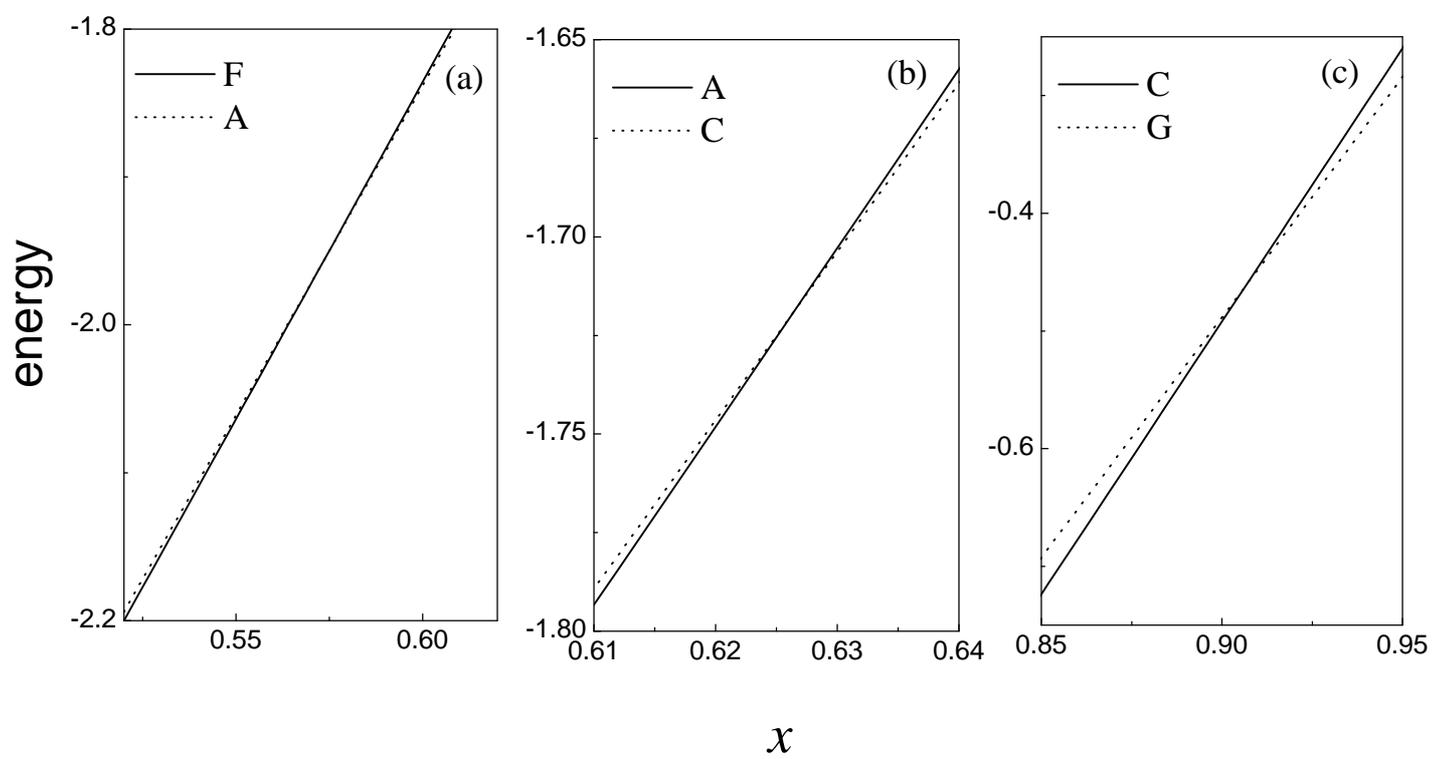
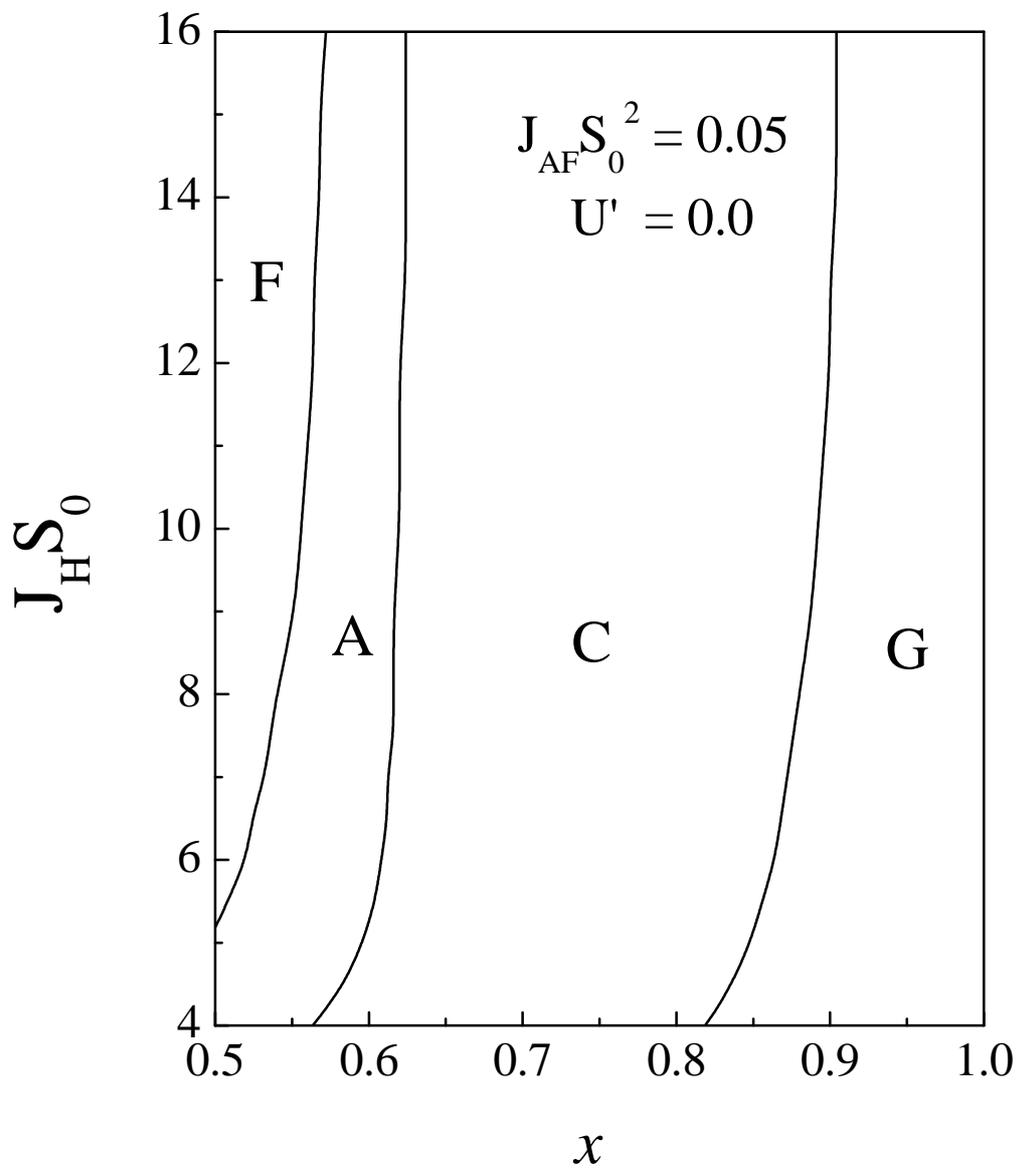
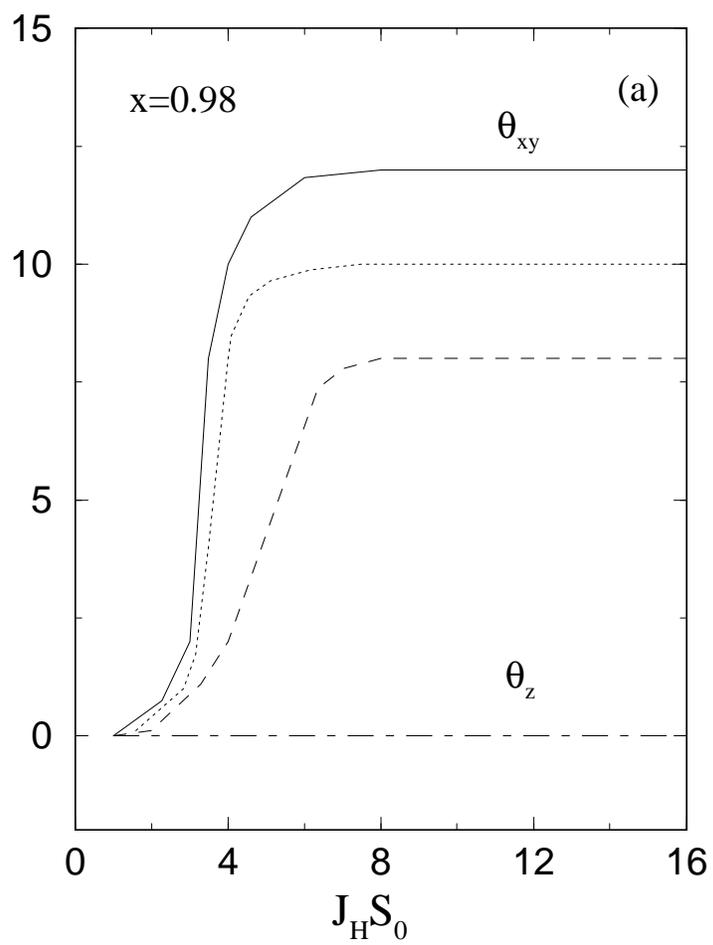
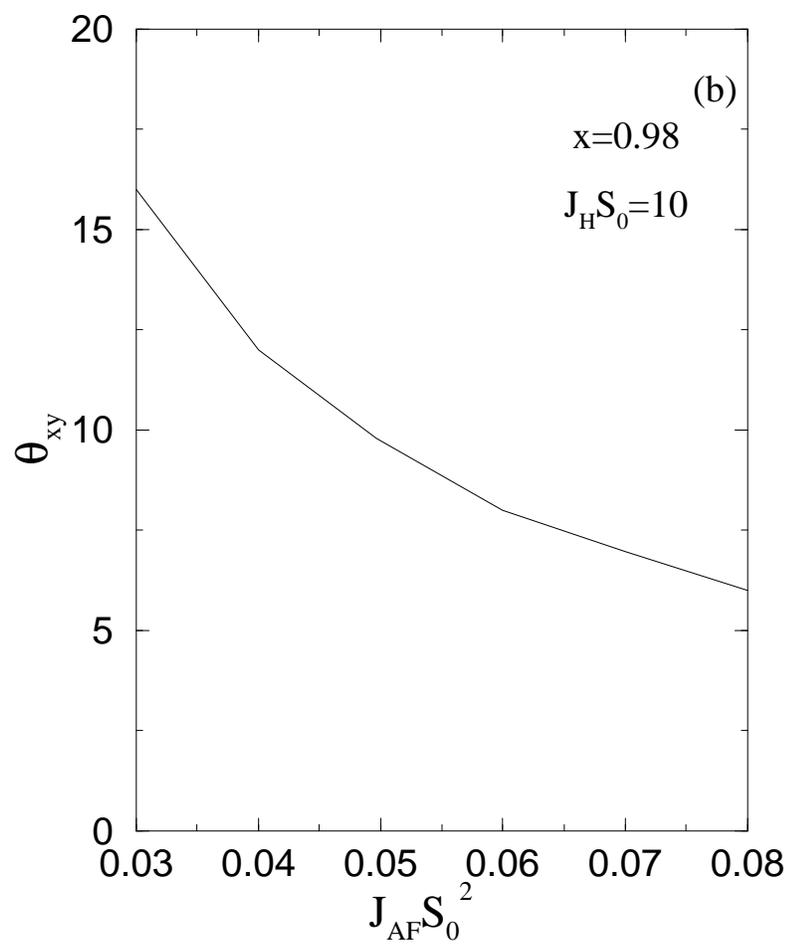
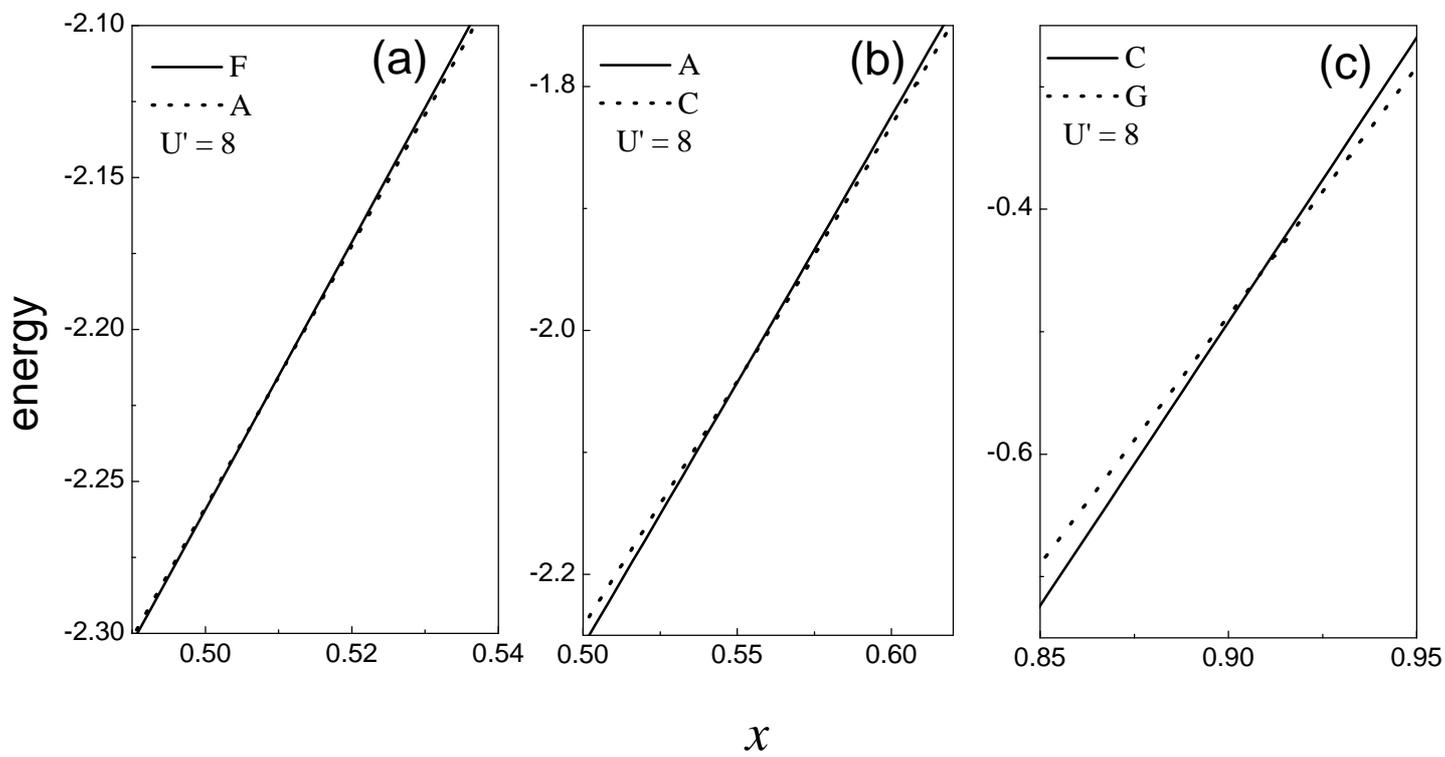
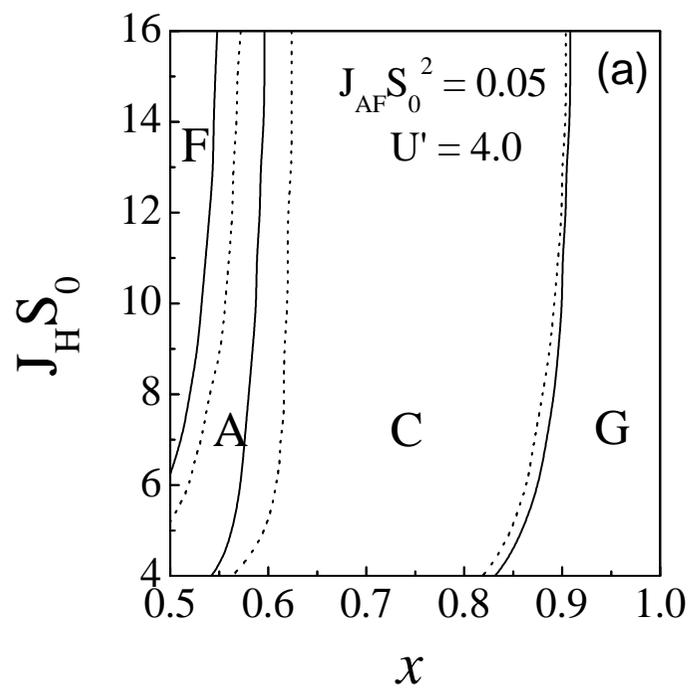
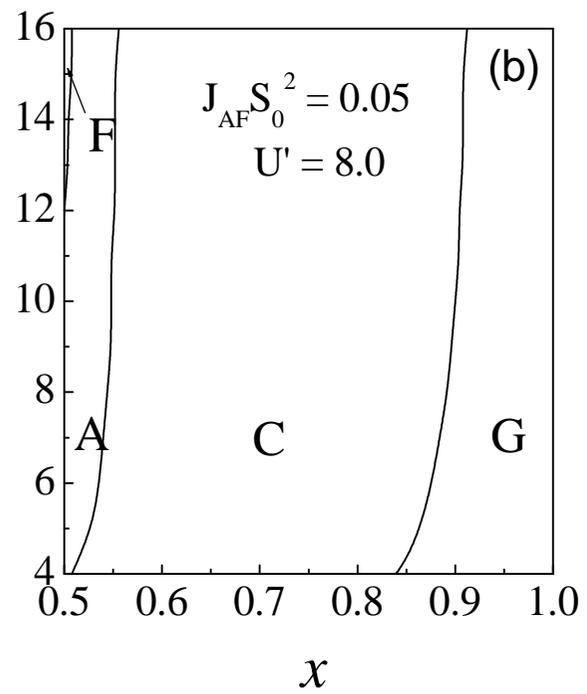
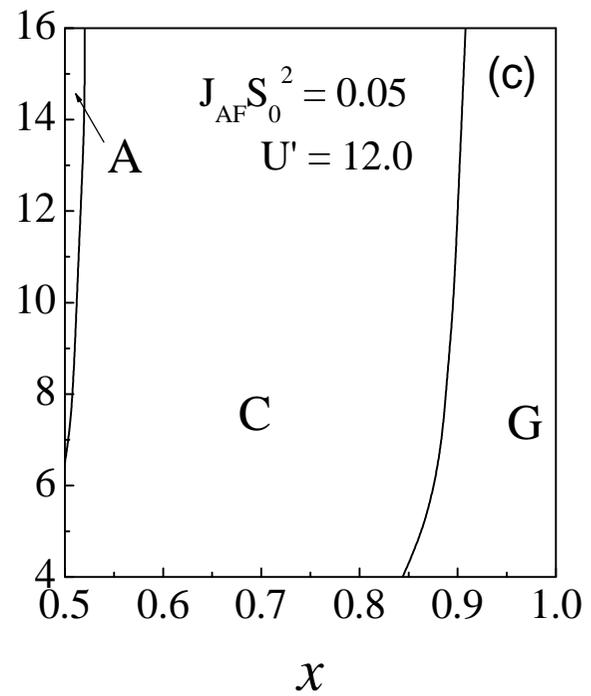
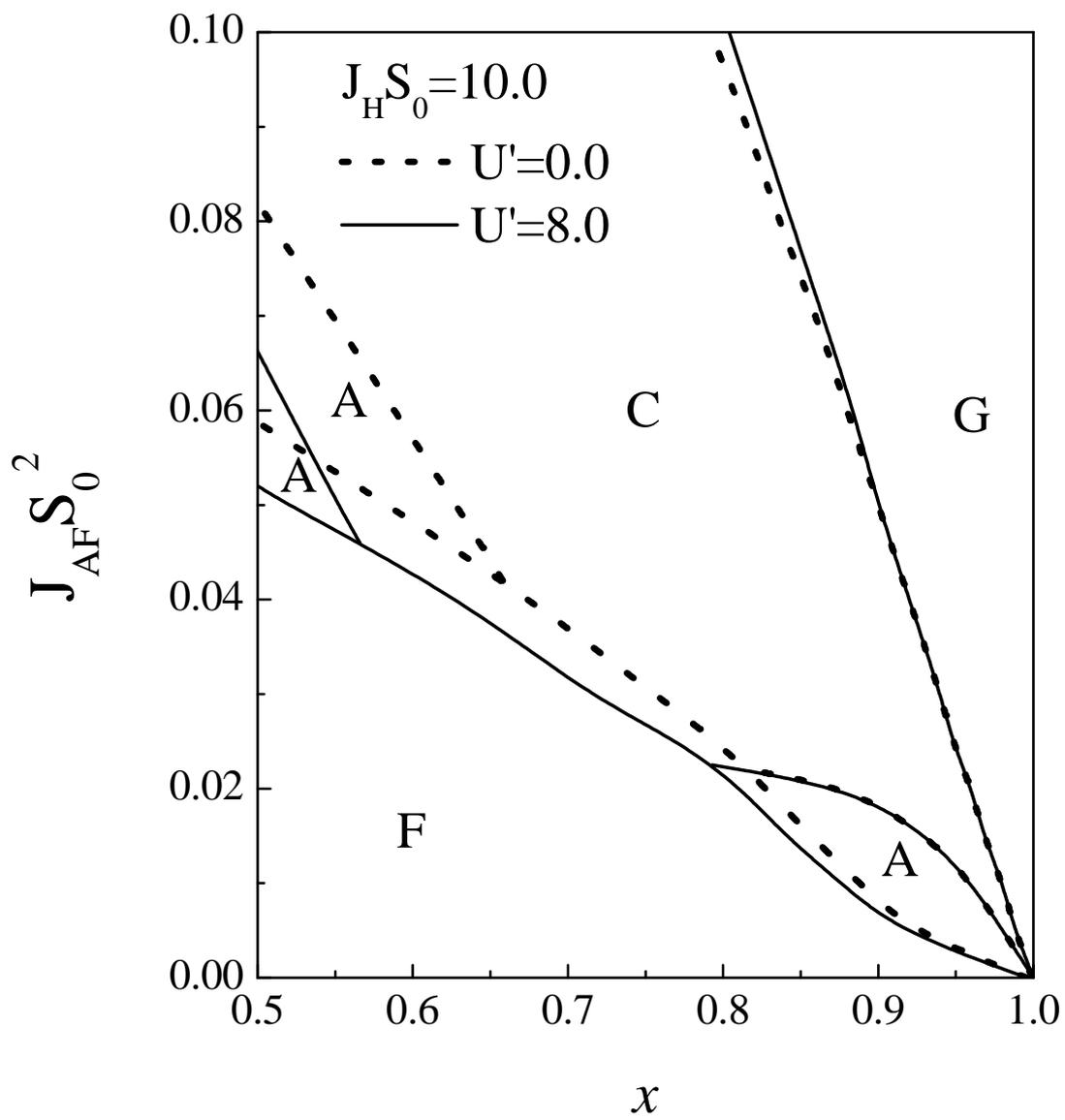
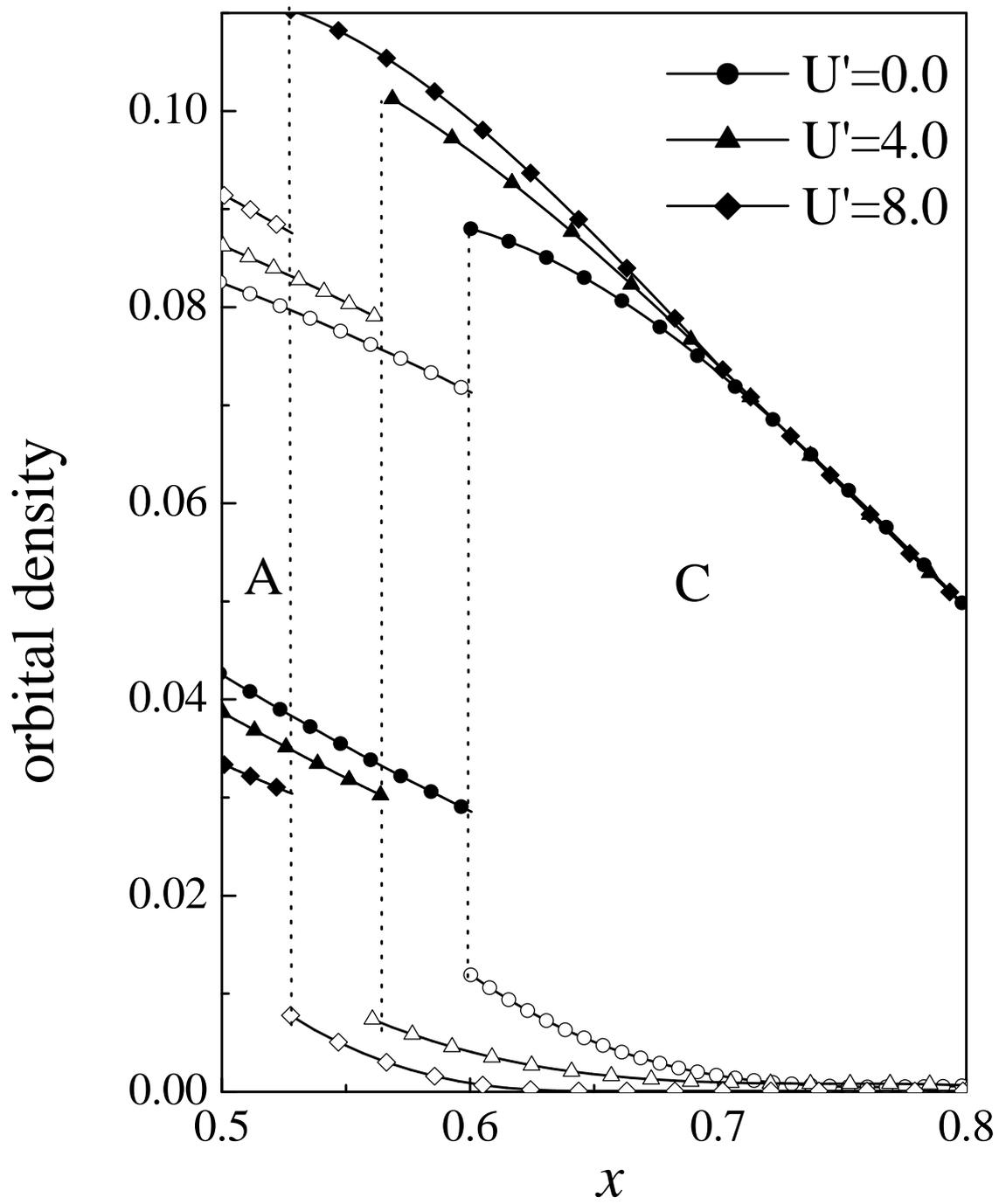
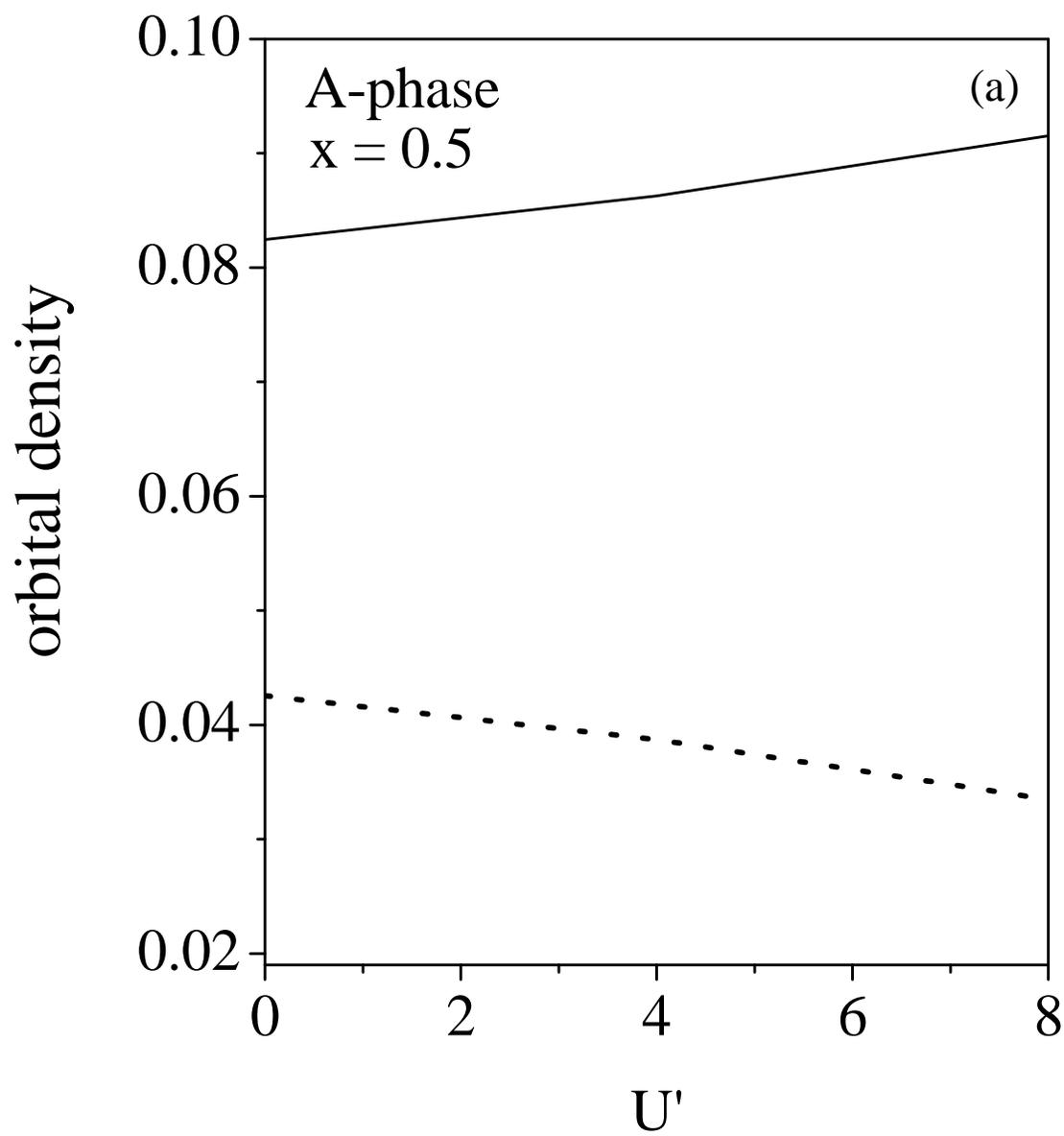
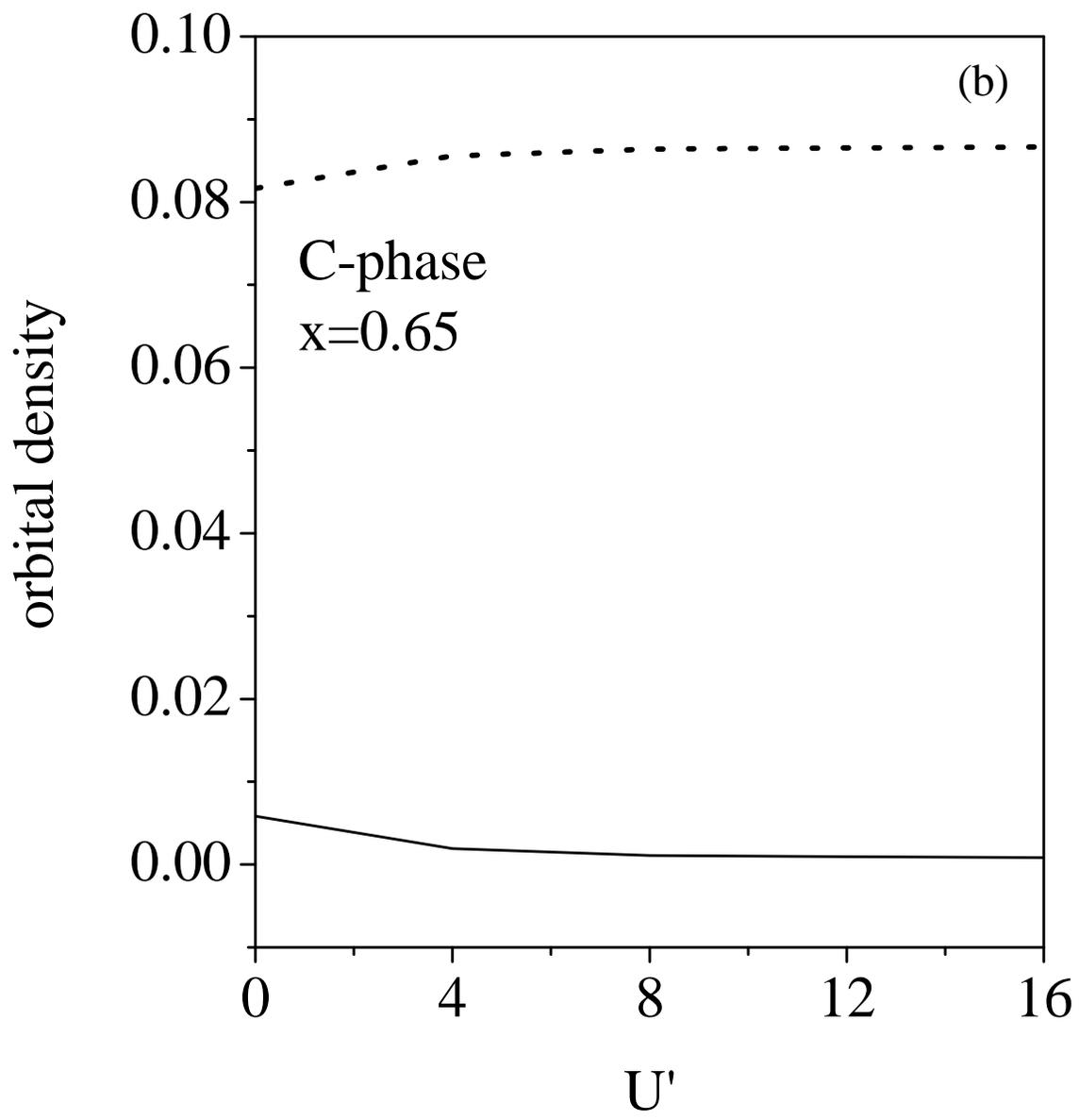
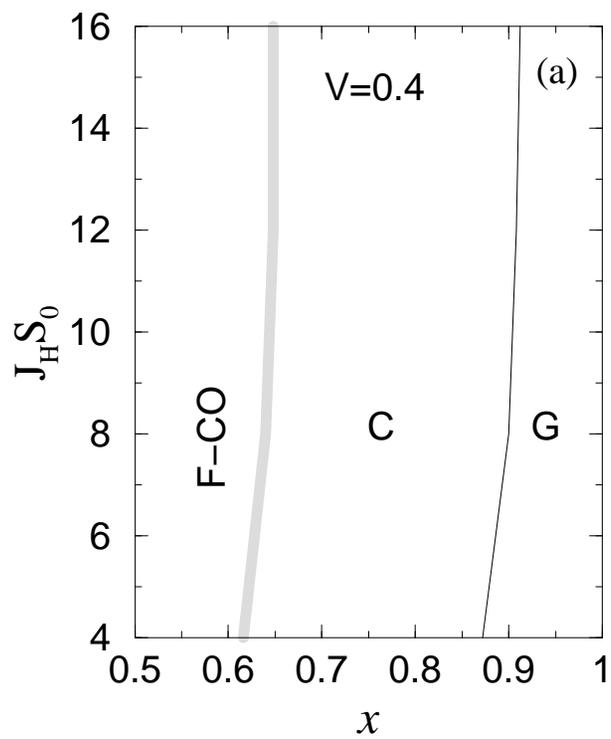
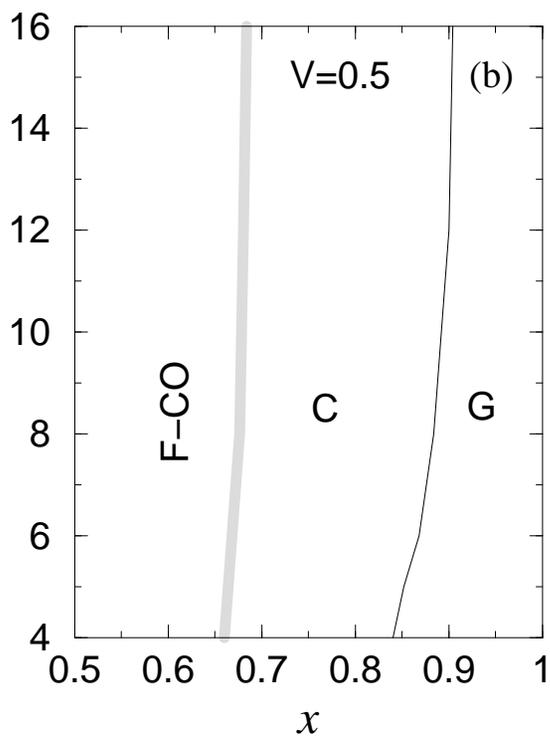
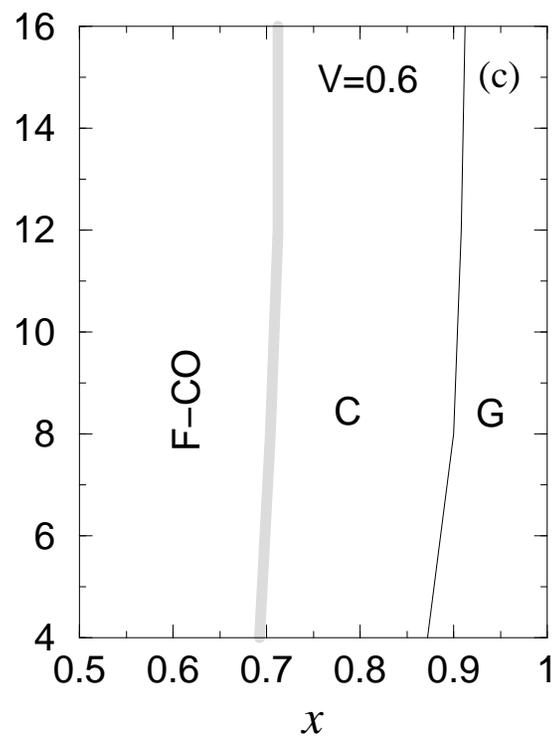
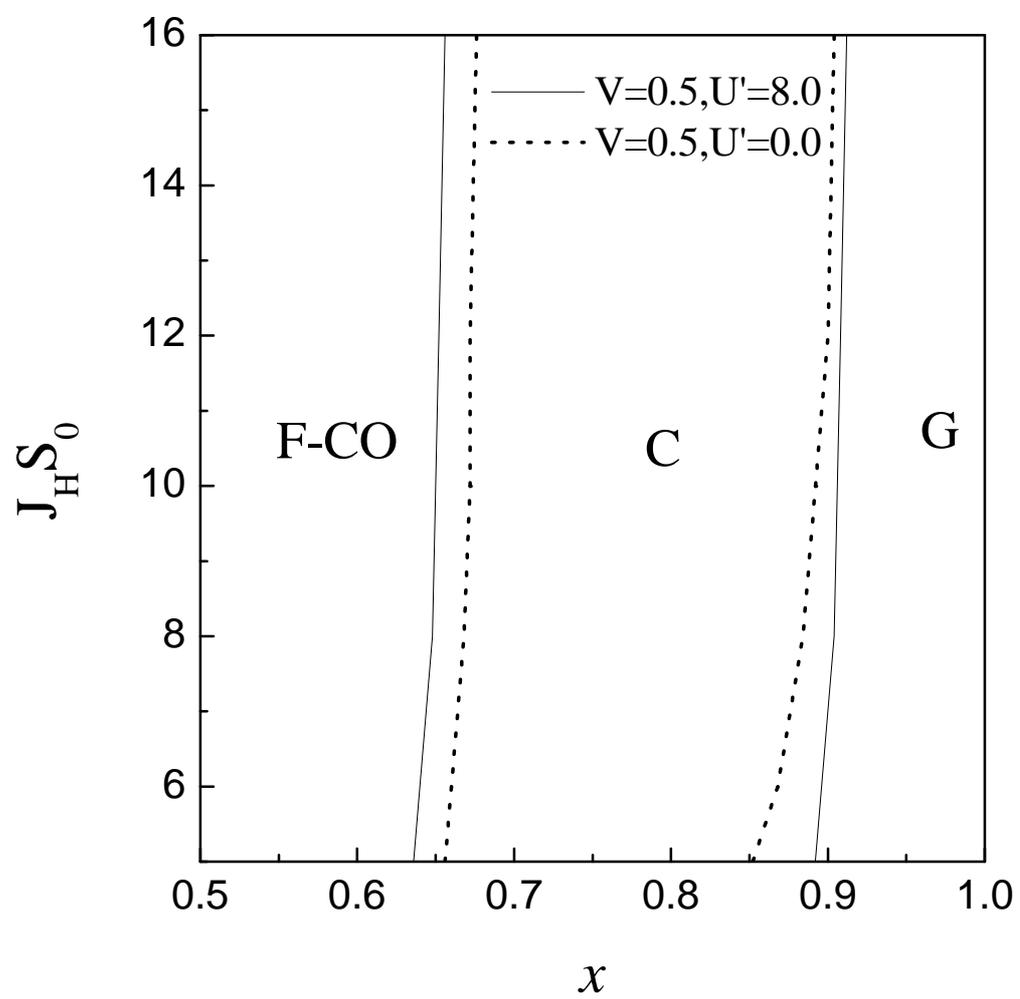
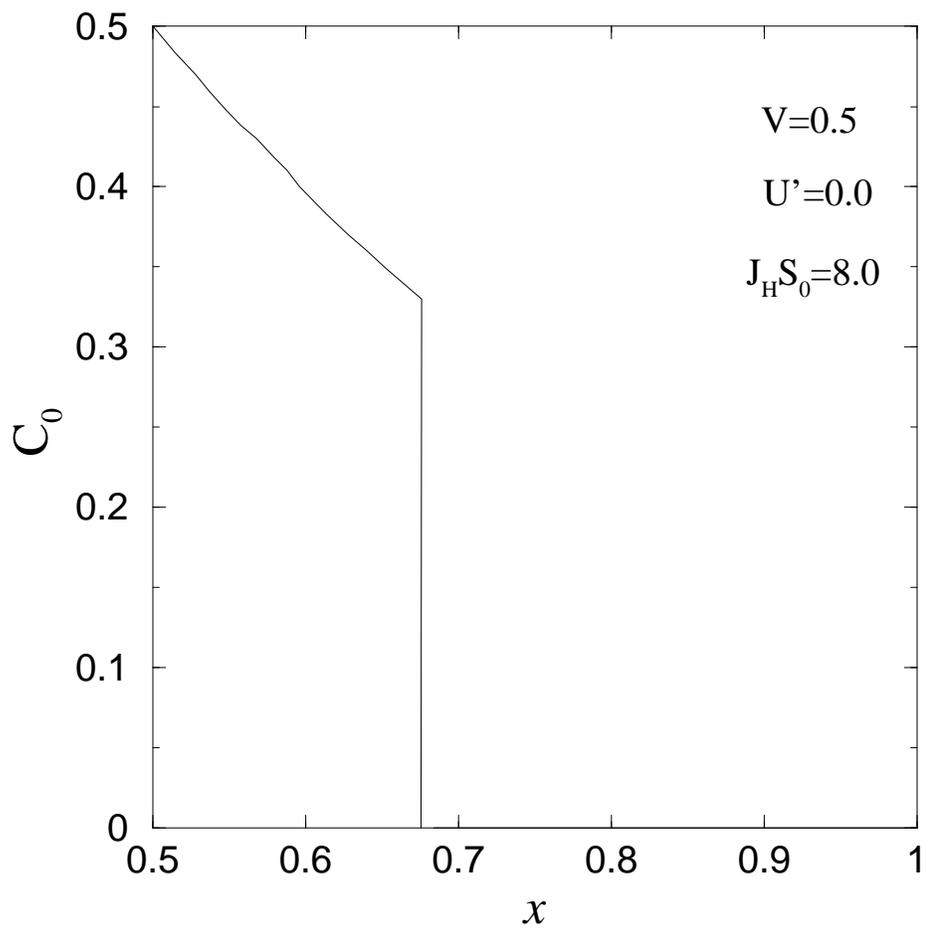
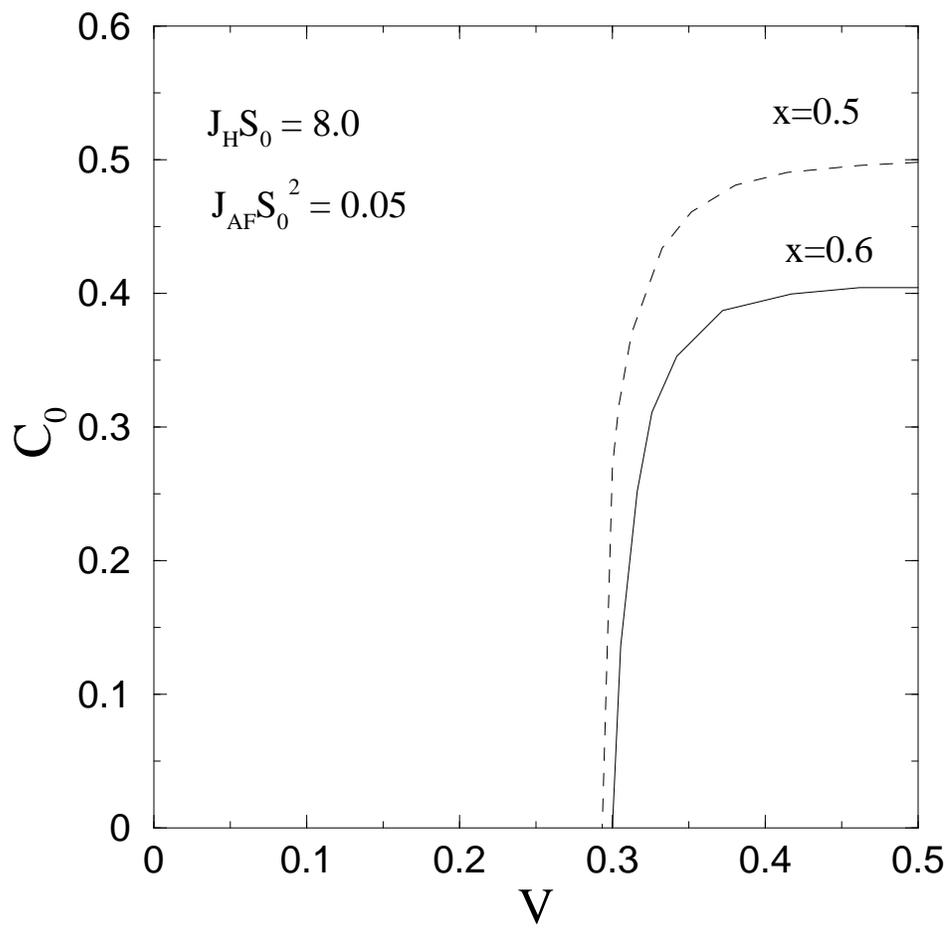
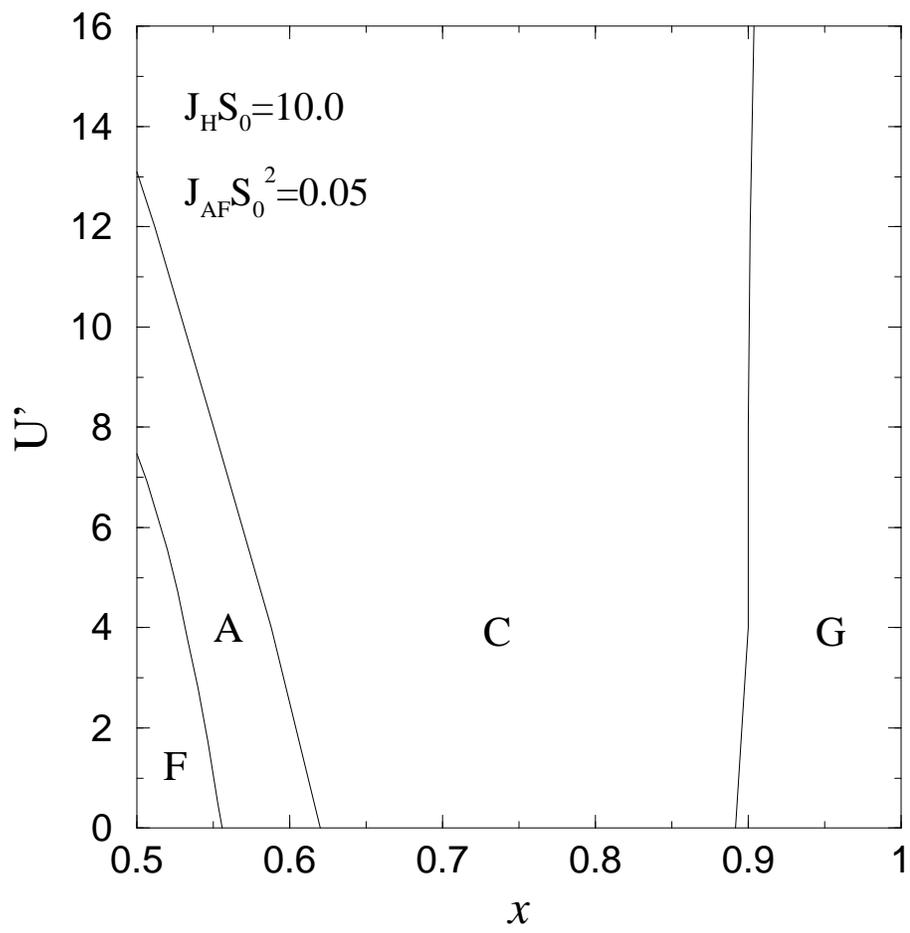
\end{document}